\documentclass[12pt,preprint]{aastex}
\usepackage{epstopdf}
\usepackage{subfigure}

\shorttitle{Accretion Shocks in Clusters of Galaxies}
\shortauthors{S.~M.~Molnar, et al.}

\newcommand{\simless} 
     {\ensuremath{\lower 3pt\hbox{$\rlap{\raise5pt\hbox{$\char'074$}}\mathchar"7218$}}}
\newcommand{\simgreat}
     {\ensuremath{\lower 3pt\hbox{$\rlap{\raise5pt\hbox{$\char'076$}}\mathchar"7218$}}}

\newcommand{\LCDM}{{$\Lambda$CDM}}

\newcommand{\tmk}{{\sc Tillamook}}
\newcommand{\GADGET}{{\sc GADGET-2}}

\newcommand{\ENZO}{{\sc ENZO}}

\newcommand{\HEALPIX}{{\sc HEALPix}}
\newcommand{\ROSAT}{{\it ROSAT}}
\newcommand{\WMAP}{{\it WMAP}}
\newcommand{\ALMA}{{\it ALMA}}
\newcommand{\MIRIAD}{{\it MIRIAD}}
\newcommand{\AIPS}{{\it AIPS}}

\newcommand{\rmsub}[1]{\ensuremath{_{\rm #1}}}
\newcommand{\MSUN}{{\ensuremath{\mbox{\rm M}_{\odot}}}}
\newcommand{\RVIR}{\ensuremath{R\rmsub{vir}}}
\newcommand{\MVIR}{{\ensuremath{M\rmsub{vir}}}}
\newcommand{\VRAD}{{\ensuremath{V\rmsub{rad}}}}
\newcommand{\RTA}{{\ensuremath{R\rmsub{ta}}}}
\newcommand{\XMAX}{{\ensuremath{x\rmsub{max}}}}

\begin{document}

\title{Accretion Shocks in Clusters of Galaxies and their SZ Signature from Cosmological Simulations}

\author{
Sandor M. Molnar\altaffilmark{1}, Nathan~Hearn\altaffilmark{2}, Zolt\'an~Haiman\altaffilmark{3}, 
  Greg~Bryan\altaffilmark{3}, \\ 
  August~E.~Evrard\altaffilmark{4}, and George~Lake\altaffilmark{5}
}

\begin{abstract}
Cold dark matter (CDM) hierarchical structure formation models predict the existence of large-scale
accretion shocks between the virial and turnaround radii of clusters of galaxies. 
\citet{Kocset05} suggest that the Sunyaev-Zel'dovich (SZ) signal associated with   
such shocks might be observable with the next generation radio interferometer, \ALMA. 
We study the three--dimensional distribution of accretion shocks around individual clusters of galaxies 
drawn from adaptive mesh refinement (AMR) and smoothed particle hydrodynamics (SPH) 
simulations of \LCDM\ (dark energy dominated CDM) models.
In relaxed clusters, we find two distinct sets of shocks. One set (``virial shocks''), 
with Mach numbers of 2.5--4, is located at radii 0.9-1.3\RVIR, where \RVIR\ is the spherical 
infall estimate of the virial radius, covering about 40-50\%
of the total surface area around clusters at these radii.  
Another set of stronger shocks (``external shocks'') is located farther out, at about 3\RVIR, with large 
Mach numbers ($\approx$100), covering about 40-60\% of the surface area.  
We simulate SZ surface brightness maps of relaxed massive galaxy clusters 
drawn from high resolution AMR runs, and conclude that \ALMA\ should be capable 
of detecting the virial shocks in massive clusters of galaxies. More simulations are needed to 
improve estimates of astrophysical noise and to determine optimal observational strategies.
\end{abstract}


\altaffiltext{1}{Institute of Astronomy and Astrophysics, Academia Sinica, 
                      P.O. Box 23-141, Taipei 106, Taiwan, Republic of China;
                      sandor@asiaa.sinica.edu.tw}
\altaffiltext{2}{ASC/Alliances Center for Astrophysical Thermonuclear Flashes, 
                      University of Chicago, Chicago, IL 60637 USA; 
                      Present address: National Center for Atmospheric Research, PO Box 3000, Boulder, CO 80503 USA; 
                      nhearn@ucar.edu}

\altaffiltext{3}{Department of Astronomy, Columbia University, 550 West 120th Street, 
                     New York, NY 10027;\\zoltan@astro.columbia.edu}

\altaffiltext{4}{Departments of Physics and Astronomy, University of Michigan, 
                      Ann Arbor, MI 48109-1040; evrard@umich.edu}

\altaffiltext{5}{Institute of Theoretical Physics, University of Zurich,
                     190 Winterthurerstrasse, Zurich, CH-8057, Switzerland; 
                      lake@physik.unizh.ch}

% % % % % % % % % % % % % % % % % % % % % % % % % % % % % % % % % % % % % % % % % % % % % % % % %
\section{Introduction}
\label{S:Intro}

According to our current structure formation scenarios, the cold dark 
matter (CDM) models, structures in the Universe form via gravitational instability.
When a mass overdensity reaches a critical threshold, the expansion of the 
density fluctuation eventually stops and the region collapses.
On large scales (\simgreat$\,$Mpc), the cooling time is much longer than the 
gravitational dynamical time, and the super--sonic collapse of gas in overdense regions 
into sheets, filaments and halos is halted by large--scale accretion shocks. 
These shocks in the low-density intergalactic gas convert the kinetic energy of
the collapsing gas into thermal energy, and are responsible for heating the gas. 
Accretion shocks are also important for generating high energy photons 
and cosmic rays (CR) via diffusive first order Fermi acceleration of charged particles
(e.g., \citealt{Miniet01a,Miniet01b,GaBl03,Mini03,Ryu_et03,Kanget07,Pfroet07,Skilet08} 
and references therein).

Semi-analytic solutions exist only for the one-dimensional accretion problem. 
\cite {FillGold1984} derived self-similar solutions for collisionless matter perturbations 
in an expanding Einstein--de Sitter universe with planar, cylindrical and spherical geometry.
Assuming spherical geometry, self-similar solutions for positive density perturbations of 
collisionless and collisional matter in an Einstein--de Sitter universe (ignoring radiation transfer, 
heat conduction and magnetic fields) were derived by Bertschinger (1985).
Semi-analytic models for large-scale structure shocks produced by spherical density
perturbations were derived by \cite{FuLo04}. 
Barkana (2004) studied the physical properties of the infalling gas 
using a spherical model that started with different initial density distributions. 
Three--dimensional (3D) cosmological N--body and hydrodynamical simulations, based on CDM models, 
confirmed the existence of large-scale accretion shocks at about the virial radius, \RVIR, 
often referred to as ``virial shocks'', around individual collapsed and 
bound structures, i.e. massive clusters of galaxies
(e.g., Evrard 1990; Bryan \& Norman 1998; Keshet et al. 2003).

Other studies based on cosmological simulations have focused on
the global properties of structure formation shocks on very large scales 
\citep{Miniet00,Ryu_et03,Pfroet06,Kanget07, Skilet08}.
For example, shocks generated by structure formation were tracked and studied in detail 
by Miniati et al. (2000) using Eulerian cosmological simulations.
They have found that the spatial structure of shocks is very complex,
multiply connected, and that shocks plunge deep into clusters of galaxies along overdense filaments. 
Shocks can be classified based on their location and/or on the physical mechanism which drives them. 
Based on their location, shocks can be classified as external and internal shocks \citep{Miniet00}.
Large-scale, external shocks form where pristine, previously un-shocked, gas is falling onto sheets, 
filaments, and halos, due to gravitational instability. External shocks are accretion shocks surrounding voids.
Internal shocks form in gas that has already been shock-heated, 
and can be further divided, based on the driving physical mechanism \citep{Ryu_et03}, 
into the following categories:

\begin{itemize}
  \item
    Internal accretion shocks, due to coherent continued infall of diffuse uncollapsed gas, following an external shock
  \item
    Merger shocks, which occur when bound or collapsing objects 
                  (such as halos, filaments, and sheets) collide
  \item    Flow shocks, which are generated within collapsed objects by internal, in general non--radial, bulk flows
   (which are not necessary gravitationally driven)
\end{itemize}

The global energetics of external and internal cosmological shocks, focusing on their role 
in accelerating nonthermal, cosmic-ray (CR) particles, were studied in detail by \cite{Ryu_et03}
using Eulerian simulations and by \cite{Pfroet06} 
using the smoothed particle hydrodynamics (SPH) code \GADGET\ \citep{Spri05}.
Using self-similar solutions, Abadi, Bower\& Navarro (2000) assessed the reliability of 
SPH cosmological numerical simulations.
They concluded that SPH simulations obtain the physical quantities describing
the collapse reliably within the virial radius. However, due to the low-density in the
pre-shock region, the smoothing length becomes large, and the pre-shock region 
is smoothed out by about 2\RVIR.
The most recent SPH simulations perform much better in the low-density regions due to 
their higher resolution.
Shock acceleration and $\gamma$-ray production were also studied by \cite{Mini02},
\cite{Keshet03}, and \cite{Kanget07}.
\cite{Skilet08} studied cosmological shocks and shock acceleration using the Eulerian code \ENZO\
\citep{BryaNorm98,OSheet04}.
Skillman et al. found that merger and internal flow shocks have Mach numbers less than about 5, and
accretion shocks into clusters of galaxies have large Mach numbers, between about 20 and 300
(they did not distinguish between internal and external shocks).
Our work here is similar, but there are two major differences:
(i) rather than using a large simulation box to compute global statistics, we utilize
a discrete sample of well--resolved individual clusters, and (ii) we focus on the
detectability of accretion shocks near the virial radius through the Sunyaev-Zel'dovich (SZ) effect
(Sunyaev \& Zel'dovich 1980; for recent reviews see Rephaeli 1995; Birkinshaw 1999;
Carlstrom, Holder \& Reese 2002).

Observational confirmation of large--scale shocks would be an important test for 
our structure formation scenarios. 
However, in general, it is difficult to detect such shocks, since the intra-cluster gas 
density is very low near, and beyond, the virial radius.
Evidence for large--scale gas infall has been claimed in the Lyman $\alpha$
spectra of some high-redshift quasars 
(see, e.g., Dijkstra, Haiman \& Spaans 2006 and references therein), 
and internal shocks are thought to be responsible for radio relics observed at smaller radii in 
nearby galaxy clusters (Ensslin et al. 1998; Ensslin 2002).
However, direct observational confirmation of large--scale accretion shocks, produced by gravitationally 
driven structure formation, at the virial radius and beyond, does not exist as of today.
A feasibility study for observations of such shocks around clusters of galaxies,
using their SZ signature, with the
next generation high-spatial-resolution, high-sensitivity interferometer, the Atacama Large
Millimeter Array (\ALMA)\footnote{See http://www.alma.nrao.edu.}, 
was carried out by Kocsis, Haiman \& Frei (2005).
The advantage of using SZ, instead of X-ray, observations is that 
the SZ signal is proportional to the gas density, while the X--ray flux is proportional to
the square of the density, and therefore
the SZ surface brightness drops considerably less rapidly with radius.
Assuming spherical models for the cluster gas and temperature profiles, 
\citet{Kocset05} concluded that strong shocks around the virial radii of clusters of 
galaxies should be detectable with high significance using \ALMA\ mosaic observations.

In this paper, motivated by the proposal of \citet{Kocset05}, we report a study of 
non-spherical accretion to quantify the 3D properties of accretion shocks 
around massive clusters of galaxies.
We have used two sets of clusters, one drawn from adaptive mesh refinement (AMR) 
and another drawn from SPH simulations of \LCDM\ models. 
First, we describe the cluster samples, the methods used to locate and analyze 
accretion shocks, and the properties of the shocks found in the AMR (\S\ref{S:AMR}) and the
SPH (\S\ref{S:SPH}) simulations. 
In the discussion section (\S\ref{S:Discussion}), first we study the feasibility of observing 
shocks in clusters of galaxies with the future high resolution radio interferometer, \ALMA\ 
using their SZ signature. 
We then discuss the consequences of our results for the interpretation of large scale 
X-ray and SZ signals associated with clusters of galaxies. 
Finally in this section, we provide a comparison of our results with those from other studies, 
as well as between our own results from AMR versus SPH simulations.
We briefly summarize our main results and offer our conclusions in \S\ref{S:Conclusion}.

% % % % % % % % % % % % % % % % % % % % % % % % % % % % % % % % % % % % % % % % % % % % % % % % %
\section{Accretion Shocks in AMR Simulations}
\label{S:AMR}

% % % % % % % % % % % % % % % % % % % % % % % % % % % % % % % % % % % % % % % % % % % % % % % % %
\subsection{Locating and Analyzing Accretion Shocks in AMR Clusters}
\label{SS:AMRShocks}

We use a sample of 10 clusters of galaxies drawn from AMR simulations,
each located at redshift $z=0$. We refer to them as AMRCL1 through AMRCL10, sorted by 
their virial mass 
(i.e. the total gravitational mass within the virial radius, \RVIR, defined as in \citealt{BryaNorm98}), 
AMRCL1 being the most massive.
The AMR simulations were performed with the cosmological code \ENZO\
 assuming a spatially flat \LCDM\ cosmology with 
($\Omega_m$, $\Omega_\Lambda$, $\Omega_b$, h, $\sigma_8$) = (0.3, 0.7, 0.047, 0.7, 0.92),
where the Hubble constant is $H_0$ = 100 h km s$^{-1}$ Mpc$^{-1}$,
and $\sigma_8$ is the power spectrum normalization on 8 h$^{-1}$ Mpc scales. 
This cosmological model is close to the \WMAP\ 5 year results, except the simulations
used a somewhat larger value of $\sigma_8$ \citep{Dunk08}.
The box size of the original simulation is 300 h$^{-1}$ Mpc.
The AMR simulations were adiabatic (in the sense that no radiative heating, cooling, 
star formation, or feedback were taken into account).
The clusters of galaxies in our sample were re-simulated with high resolution 
using the same  technique as described in \cite{YounBria07}.
The virial masses of the clusters in our AMR sample are in the range of 
9.1$\times 10^{14}$ \MSUN\ - 2.3$\times 10^{15}$ \MSUN; in Table~\ref{T:Table1}, we list  the
masses, radii, and dynamical state (see below) of each cluster.

\begin{figure}
\centerline{
\plotone{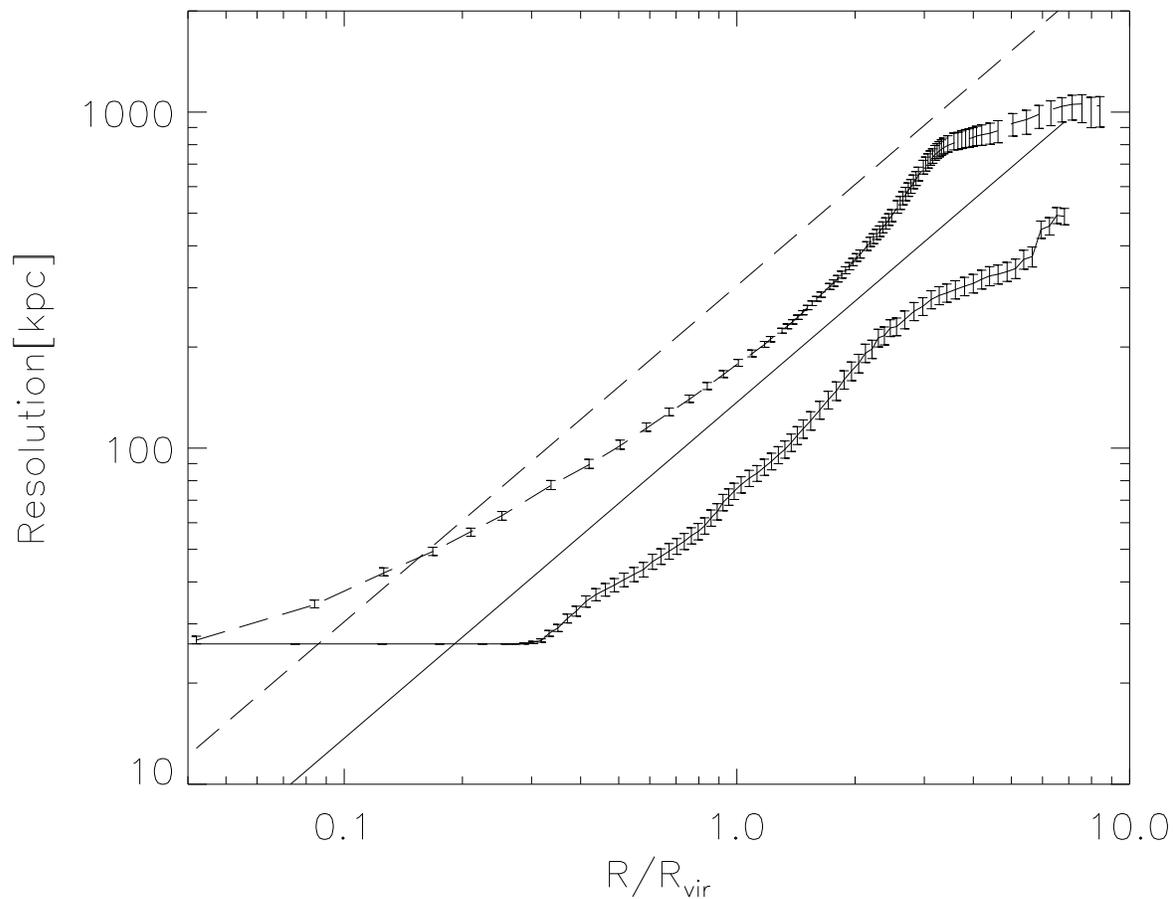}
}
\caption{
 Spherically averaged, position--dependent spatial resolution in the relaxed clusters AMRCL9 and SPHCL4
 (points with error bars connected with solid and dashed curves).
 The straight solid and dashed lines show the linear sizes of the angular \HEALPIX\ 
 pixels in physical units (kpc) used to analyze AMR and SPH clusters.
\label{F:Fig1}
}
\end{figure}

We are interested in accretion shocks around each cluster of galaxies;
therefore, we use spherical coordinates with the origin always placed at the 
center of mass of each cluster.
We sample physical variables from the simulation output at spatial positions 
$\vec{x} = (R, P)$, described by the radial distance $R$ from the cluster center, and by 
the 2D angular position specified by a pixel number $P$, based on the \HEALPIX\ 
pixelization scheme with 3072 pixels \citep{Gorset05}.
This coordinate system provides a convenient way of describing physical
variables around collapsed objects. 
A 3D volume element in our coordinate system has the shape of a two-dimensional (2D) 
\HEALPIX\ pixel with a thickness of the corresponding radial bin, $\Delta R$.
The physical size of the 3D volume elements are larger at larger radii, but, in our case, 
that is desired because the resolution of the simulations also decreases with 
distance from the cluster center (except for substructures, which are not 
important for our analysis). 
As an example, we show the resolution (spherically averaged AMR cell size)
as a function of radius in a typical relaxed cluster, AMRCL9 
(\MVIR =1.1$\times 10^{15}$ \MSUN), in Figure~\ref{F:Fig1} 
(points with error bars connected with solid line). 
The error bars represent the dispersion due to spherical averaging.
We estimate the dispersion by calculating the standard deviations separately 
for data points that are larger and smaller than the mean (upper and lower error bars).
With the choice of 3072 pixels, the \HEALPIX\
pixel sizes as a function of radius track the resolution of the simulations well 
(within a factor of two; compare the solid line with error bars and the straight
solid line in Figure~\ref{F:Fig1}).
The median resolution (AMR cell size) of the high resolution simulations 
at $R/R_{vir}$ = 0.1, 1 and 4 (where $R$ is the distance from the center of the cluster)
was about 25, 80 and 300 kpc, respectively.

We carry out our analysis of shocks in all clusters, but we quote results
for relaxed clusters separately.
We define relaxed clusters as clusters without a sign of a recent merger. 
Recent mergers are identified by visual inspection of the 3D 
distribution of dark matter, the spherically averaged density, 
and temperature distributions within the virial radius, \RVIR.
Head-on mergers in a stage close to pericentric passage
would not be identified by this method; however, we ignore this possibility, since
such events are exceedingly rare, and are not expected to occur in our small sample.
In total, we find three relaxed clusters in our AMR sample.

\begin{figure}
\centerline{
\plotone{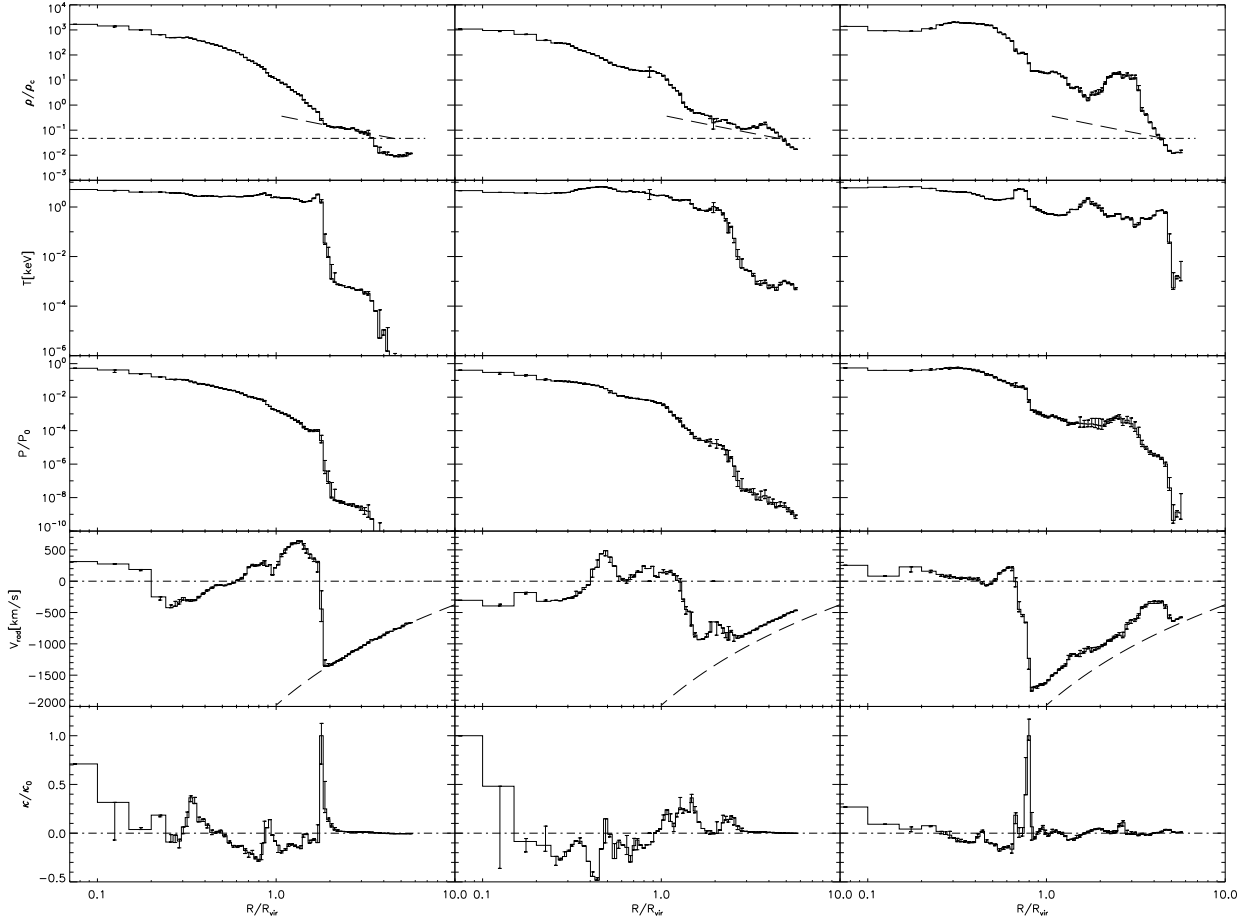}
}
\caption{
 Radial profiles of gas density, temperature, pressure,
 radial velocity, and velocity field convergence, from top to bottom,
 as a function of distance (in units of \RVIR) from the center of the relaxed AMR cluster AMRCL9.  
 The left and middle panels show two individual sight--lines that avoid overdense filaments, whereas the right
 panels represent a direction toward a filament. 
 The average universal background baryon density is plotted as a horizontal 
 dot-dashed line on the density plots.
 The dashed curves on the density and radial velocity plots show
 the self-similar collapse solution of Bertschinger (1985).
\label{F:Fig2}
}
\end{figure}

In AMR simulations, the initial cells are adaptively divided into sub-cells, and the 
physical parameters are represented by values assigned to the center of the cells.
We use a piecewise constant interpolation scheme to determine the physical variables 
at the center of our 3D volume elements.
Shocks in our AMR sample of clusters of galaxies are identified based on regions of large 
compression, {\it i.e.}, where the velocity convergence, $\kappa = -\nabla \cdot \vec{v}$, is large.  
The large peak in the convergence accurately determines the shock position
as demonstrated in Figure~\ref{F:Fig2}. 
We use this figure to illustrate the features of different shocks as well.
In Figure~\ref{F:Fig2} 
we show the distribution of physical variables as a function of radius in three directions 
(fixed pixels: $P = A$, $B$ and $C$; right, middle and left panels) in a relaxed AMR cluster, AMRCL9. 
Similar features can be seen on all other relaxed clusters.
We plot the gas density (in units of the critical density, $\rho_c$),
temperature (in keV), pressure (in units of the central pressure, $P_0$), radial velocity (in km/s), 
and convergence profiles (in units of $\kappa_0 = \mbox{\rm Max} \{ \kappa \}$) with error bars. 
The error bars represent the dispersion (standard deviation) of a given parameter 
evaluated in all AMR cells located within the 3D sampling volume element.
Note, however, that these values are only rough estimates because some 
volume elements contain only a few points
(we omitted error bars for those 3D volume elements that contain only one point).
The dot-dashed lines on the density plots represent the universal background baryon 
mass density ($\Omega_b = 0.047$).
Pixels $A$ and $B$ are chosen toward directions that avoid any overdense filaments, 
Pixel $C$ points toward a filament.

The left panels (Pixel A) represent an individual sight--line from the cluster center 
to a low--density,  no--filament region with only one accretion shock.
For reference, the dashed curves on the density and radial velocity plots show the 
self-similar collapse solution of \cite{Bert85} for the pressureless  infall region.
We fit the functional form of the radial velocity, \VRAD, 
obtained by Bertschinger (Equation 2.13 of Bertschinger 1985) to the 
spherically averaged distribution of \VRAD\ in this direction
allowing the amplitude and the turn--around radius, \RTA, to change (dashed curve).
From this, remarkably good fit, extrapolating the curve out to the point where $\VRAD = 0$,
we derive $\RTA = 22 \RVIR$
(note that this large value is the turn--around radius at $z=0$ and not the turn--around radius of 
the shells presently located near the virial radius). 
We use this \RTA, and fit for the amplitude of the gas density distribution in this direction
using Equation 2.15 of \cite{Bert85} (dashed curve).
We obtain good fits for both the radial velocity and density profiles, thus we conclude that,
in the undisturbed pre-shock (infall) regions, both distributions 
are consistent with the functional forms derived by \cite{Bert85}. 
Note, that the radial velocity is not zero within the virial radius due to internal flows.
In this direction, we can identify a virial shock at 1.8\RVIR; at this radius, 
we see a large peak in the convergence associated with a significant decrease in pressure with radius, 
and the radial velocity of the gas falling toward the cluster center drops to zero.
The peak in the convergence at 0.35\RVIR\ is due to an internal flow shock.
We see no other accretion shock, therefore, in this direction, the virial shock is also an external shock.
The density, temperature, and pressure drop significantly 
between 3--4~\RVIR, well into the free collapse region, breaking the self-similarity. 
This drop is not due to a shock, since the radial velocity is continuous, still consistent with 
the self-similar solution, and there is no peak in the convergence. 
Rather, this feature is due to voids between massive clusters of galaxies;
the infall toward our cluster simply runs out of gas.

Pixel B in the middle panels of Figure~\ref{F:Fig2} shows a representative
line of sight with two accretion shocks. Most lines of sight toward low--density no--filament 
regions (80--90\%) have two or more accretion shocks.
The dashed curves on the density and radial velocity plots toward this region 
show the self-similar collapse solutions of \cite{Bert85} with parameters fixed at their
values derived from the profiles toward Pixel A.
In this direction, the viral shock is located at 1.1\RVIR, 
where there is a peak in the convergence, a drop in the pressure, and the radial infall is halted.
Outside of the virial radius, the density profile does not deviate significantly from that predicted 
by the self--similar solution of Bertschinger.
The radial velocity profile is close to the self--similar solution in the free infall region. 
At 2.5\RVIR\ the velocity of the infall becomes smaller than that of the free fall, 
there is a peak in the convergence, and a large jump in the pressure:
this is the radius where the external shock is located in this direction. 
An internal flow shock due to an outward bulk flow is located at 0.5\RVIR, 
where we see a small peak in the convergence, and a drop in the pressure.

Pixel $C$ in the right panel of Figure~\ref{F:Fig2} 
shows a representative sight-line toward a direction where accretion occurs along a
filament, and where the infalling gas 
forms a shock inside the virial radius (in this case at 0.8\RVIR).
We see multiple peaks in the convergence in this direction.
Density and temperature bumps can be seen at the location of
collapsed objects falling toward the cluster along the filaments.
The dashed curves on the density and radial velocity plots toward this region 
show the self-similar collapse solutions with parameters fixed as before.
The density is much larger than that predicted by the self similar collapse 
model (the property we use to identify filaments).
As expected, the self--similar solutions of Bertschinger are not good
descriptions of the collapse in this direction that includes a filament.

Since we expect more coherent accretion shocks to occur in the low-density diffuse gas 
(in regions where the infall is not disturbed by overdense filaments or sheets along the infall 
path), we identify and remove pixels in the directions of filaments. 
Note that because filaments bend and become more radial near the 
virial radius, and because they miss the cluster center, the projected surface area 
(number of pixels) of filaments as viewed from the cluster center will be larger than the area 
actually covered by filaments at a fixed distance 
(which is usually around one-third of total surface area; see below). 
It is well known that the morphology of the density distribution in cosmological simulations 
depends on the density threshold chosen: high density thresholds results in a set of 
collapsed halos; at lower density thresholds filamentary structures emerge, and at even lower 
density thresholds we see a network of sheets. 
Colberg, Krughoff \& Connolly (2005) identified and classified filaments in \LCDM\ cosmological 
numerical simulations using the dark matter density distribution between clusters of galaxies.
In order to identify filaments, we visually inspect the gas density distribution around each cluster
and choose the gas density threshold that results in a filamentary structure 
(typically 1-1.5 $\rho_c$).
While this threshold changes with distance from the cluster center, we verified that 
varying it within reasonable bounds (a factor of 5) does not affect our results.

\begin{figure}
\centerline{
\plotone{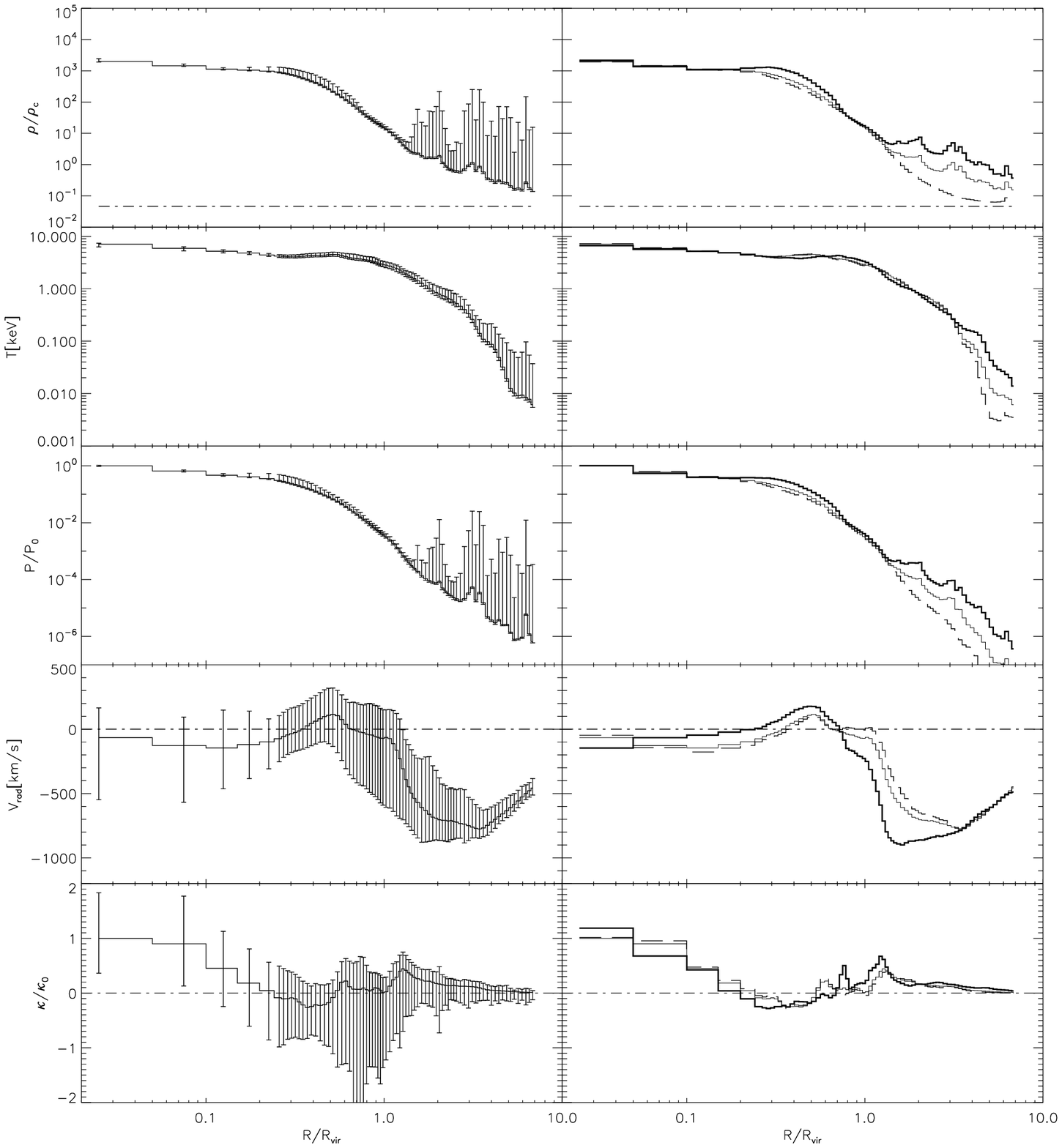}
}
\caption{
Same as Figure~\ref{F:Fig2}, except the left panels show averages over all directions (pixels),
 and the right panels show averages toward directions 
 with filaments (thick solid curves), no filaments  (dashed curves) and, for reference, 
 toward all directions (thin solid lines). 
\label{F:Fig3}
}
\end{figure}

We illustrate the results of this selection for the relaxed cluster AMRCL9 in Figure~\ref{F:Fig3} 
(other relaxed clusters show very similar characteristics). The figure shows spherically averaged profiles, 
along radial sight--lines that either explicitly avoid or explicitly include filaments.
From top to bottom, we plot profiles of the density, temperature, 
pressure, radial velocity, and convergence as a function of radius 
(we use the same units as in Figure~\ref{F:Fig2})
We show mean values over all directions (pixels) with error bars which
represent the dispersion due to spherical averaging (left panel), 
and mean values of physical quantities toward filaments (thick solid curves), 
no--filaments regions (long dashed curves), 
and in all directions (thin solid curves) for reference (right panel).
The upper and lower error bars are calculated as standard deviations for data points that are
larger and smaller than the mean. 
The average universal baryon background density, $\Omega_b$, is plotted as a dot-dashed line.
The curve representing the density distributions for filaments and no--filament regions splits 
at about 1.2\RVIR, where the radial velocity, \VRAD, reaches zero; this is, on average, 
the location of the innermost accretion shock, the ``virial'' shock, in this cluster.
The large dispersion in the density, temperature, and thus in the pressure data greater 
than the mean (upper error bars) outside of the virial shock is
caused by clumps and collapsed halos falling toward the cluster center along filaments. 
The density is much lower in the no--filament regions, and the spherically averaged
density over all directions is very bumpy in $R$ due to clumps in the filaments.
However, the spherically averaged density distribution in the low-density diffuse
gas is smooth out to large radii, and slowly drops to the background level at about 4\RVIR.
We conclude that, outside of the virial radius, spherical averaging is not a good 
approximation for the distribution of baryons.
The radial velocity average toward filaments reaches zero at about 0.6\RVIR,
much closer to the cluster center than the virial radius.

We expect coherent accretion shocks primarily in gas that falls nearly radially, unaffected
by filaments and sheets. Therefore, as a second (and last) step in locating such coherent
accretion shocks, among the regions we identified as non--filamentary based on their density 
(in our first step), we further select pixels where the infalling gas velocity is nearly radial 
($-\VRAD / V \ge 0.85$). 
This second step assures us that we isolate truly infalling 
low-density regions, as opposed to low-density regions that might be disturbed by incoherent,
non-radial motions.  In practice, we found that non--radial accretion shocks in the non--filamentary
regions are rare, and this second step provides only a small correction. 
This is reassuring, since it may be difficult to disentangle radial and non--radial shocks in observations.

The determination of Mach numbers for shocks in numerical simulations in general is not a trivial 
task (see, for example, \citealt{McCaret07,Pfroet06,Skilet08}).
However, since we found the gas velocities through the accretion shocks in the non--filamentary directions 
to be close to radial, we assume that the shock surface is tangential to the radial direction. 
At any fixed direction (pixel number), we use the Rankine-Hugoniot jump conditions 
to determine the upstream Mach number from the pressure ratio taken from the simulations:
\begin{equation}
 M_1^2 = {(\gamma + 1) P_2/P_1 + (\gamma - 1) \over  2 \gamma}
,
\end{equation}
where $P_1$ and $P_2$ are the upstream (pre-shock) and downstream (post-shock)
pressures at the position of the shock,
and $\gamma = 5/3$.
Although we assume radial shocks, our method still gives a good approximation
for the Mach numbers of shocks as long as the shock surface is not in the
radial direction (the number of pixels with the shock surface close to
radial is indeed negligible).

% % % % % % % % % % % % % % % % % % % % % % % % % % % % % % % % % % % % % % % % % % % % % % % % %
\subsection{Results: AMR Clusters}
\label{SS:AMREsults}

We next describe the properties of the shocks we identified around the
AMR clusters. As will become clear below, there are two distinct sets
of accretion shocks, which can be distinguished by their Mach numbers, and also
by their radial location.  We first discuss the ``virial'' shocks,
which are located closer in, and have low Mach numbers. 
We then discuss the ``external'' shocks, which
are located farther out, with large Mach numbers, and where infalling
gas is shocked for the first time.

% % % % % % % % % % % % % % % % % % % % % % % % % % % % % % % % % % % % % % % % % % % % % % % % %
\subsubsection{Virial Shocks}
\label{SSS:VShocks}

We fit Gaussian functions to the peaks of the convergence profile for the sight--lines in 
no--filament directions (excluding the few pixels with non--radial infall).
We determine the position of the virial shock as 
the radial position of the shock near the virial radius toward no--filament regions, where the radial infall 
is halted (the radial velocity drops to zero the first time moving toward the cluster center).
We illustrate our results for AMR clusters using the relaxed cluster AMRCL5, but other 
relaxed clusters show similar features. 
In Figures~\ref{F:Fig4}a and \ref{F:Fig4}b we show histograms of the radial position 
and Mach number of virial shocks in AMRCL5. 
These figures show that the shock position distribution in this cluster peaks at about 
${\cal R}\rmsub{vsh} = \mbox{\rm Mode} \{R\rmsub{vsh} \}= 1.1 \RVIR$, and has a width of about 
$\Delta {\cal R}\rmsub{vsh}\approx 0.5 \RVIR$; the Mach number distribution peaks at 
${\cal M}\rmsub{vsh} = \mbox{\rm Mode} \{M\rmsub{vsh} \}\approx 3.5$, 
and has a broad distribution with a width of about $\Delta {\cal M}\rmsub{vsh}\approx 3$. 
A map of the 2D angular distribution of the virial shocks in this cluster is shown in Figure~\ref{F:Fig5}a.

The parameters of the virial shocks in all of our AMR clusters are summarized in Table~\ref{T:Table1}. 
Columns $e$, $f$ and $g$ show the characteristic radial position of the virial shocks in different clusters, 
${\cal R}\rmsub{vsh} = \mbox{\rm Mode} \{ R\rmsub{vsh} \}$, the characteristic Mach numbers,  
${\cal M}\rmsub{vsh} = \mbox{\rm Mode} \{ M\rmsub{vsh} \}$, and the surface areas covered by virial shocks, ${\rm A}\rmsub{vsh}$.
The positions of the virial shocks in these clusters fall between 1\RVIR\ and 1.5\RVIR.
The characteristic Mach number of virial shocks for all clusters
in our AMR sample is about 3.
The distribution of the surface area covered by virial shocks is broad, 
falling between 18--46\% of the total surface area.
In the relaxed clusters (AMRCL2, 5, 9), virial shocks cover 37--46\% of the total area.

% % % % % % % % % % % % % % % % % % % % % % % % % % % % % % % % % % % % % % % % % % % % % % % % %
\subsubsection{External Shocks}
\label{SSS:XShocks}

External shocks have high Mach numbers due to the fact that the un-shocked gas is very cold. 
Since our high--resolution cluster simulation data are in a box cut out from a larger simulation, 
we can not determine directly, based on geometry, whether a given shock is internal or external.
Therefore, we use a two--step procedure to locate external shocks.
First, we remove shocks in filaments using the method we described in Section~\ref{SS:AMRShocks}.
Then, based on the Mach number distribution of shocks, we determine a cut--off Mach number 
for external shocks and identify external shocks as shocks with Mach number greater than this
cut--off value.

\begin{figure}
     \centering
     \subfigure[]{
          \includegraphics[width=.48\textwidth]{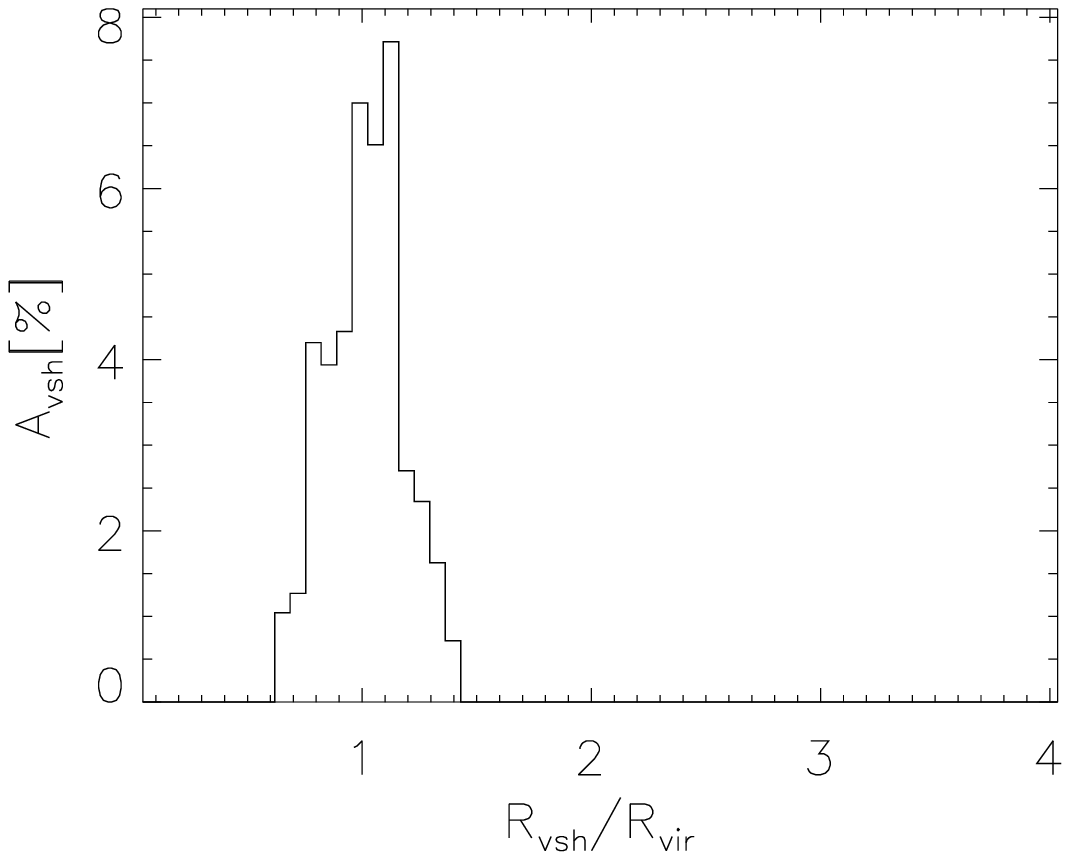}}
     \subfigure[]{
          \includegraphics[width=.48\textwidth]{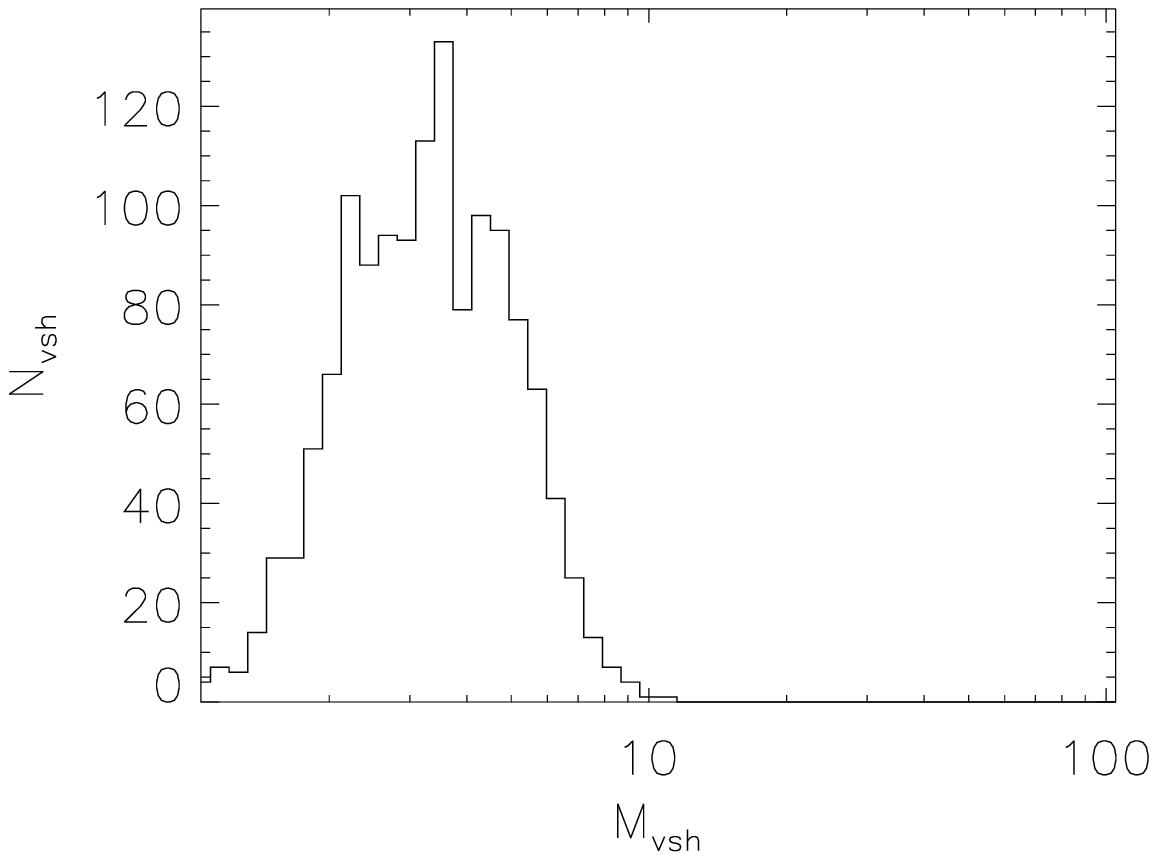}}
     \subfigure[]{
          \includegraphics[width=.48\textwidth]{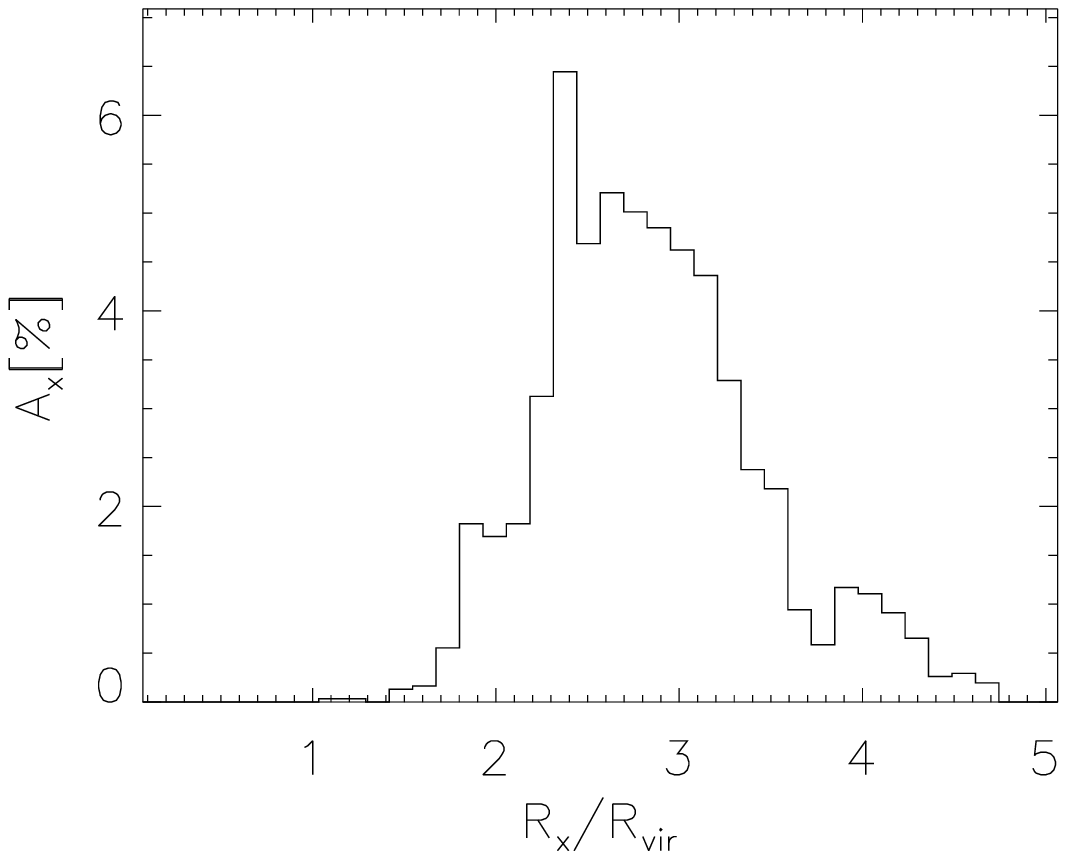}}
     \subfigure[]{
          \includegraphics[width=.48\textwidth]{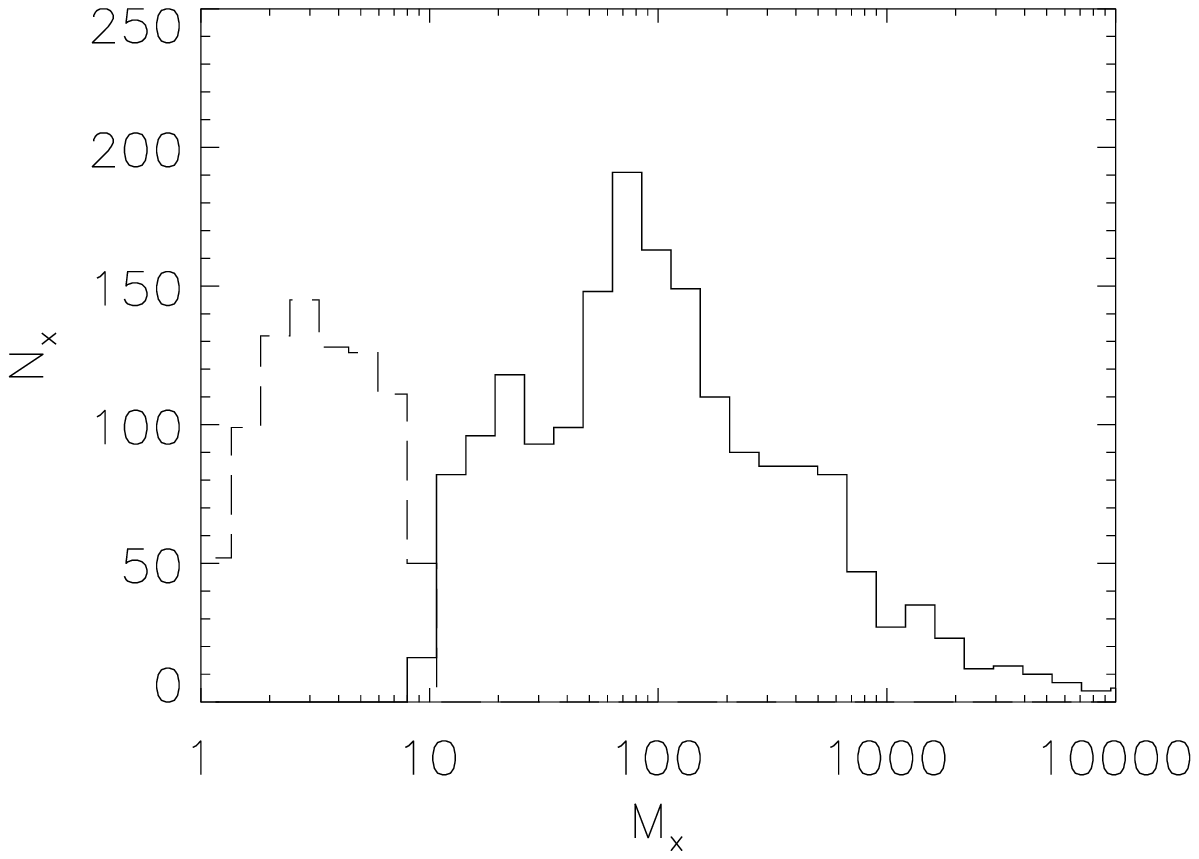}}
\caption{
         Histograms of shock positions and Mach numbers 
         for virial (a: $R\rmsub{vsh}$, and b: $M\rmsub{vsh}$), 
         and external (c: $R_x$, and d: $M_x$) shocks in the relaxed cluster AMRCL5.
         The shock--position histograms, $A_{vsh}$ and $A_x$, are normalized to show the percentage of the 
         total surface area covered by shocks in each radial bin.
         $N\rmsub{vsh}$ and $N_x$ represent the number of virial and external shocks per radial bin.
         In panel (d), we also show the Mach number distribution of the shocks we excluded
         as internal shocks (dashed line) based on our Mach
         number cut--off criterion (see text for details). 
\label{F:Fig4}
}
\end{figure}

\begin{figure}
     \centering
     \subfigure[]{
          \includegraphics[width=.9\textwidth]{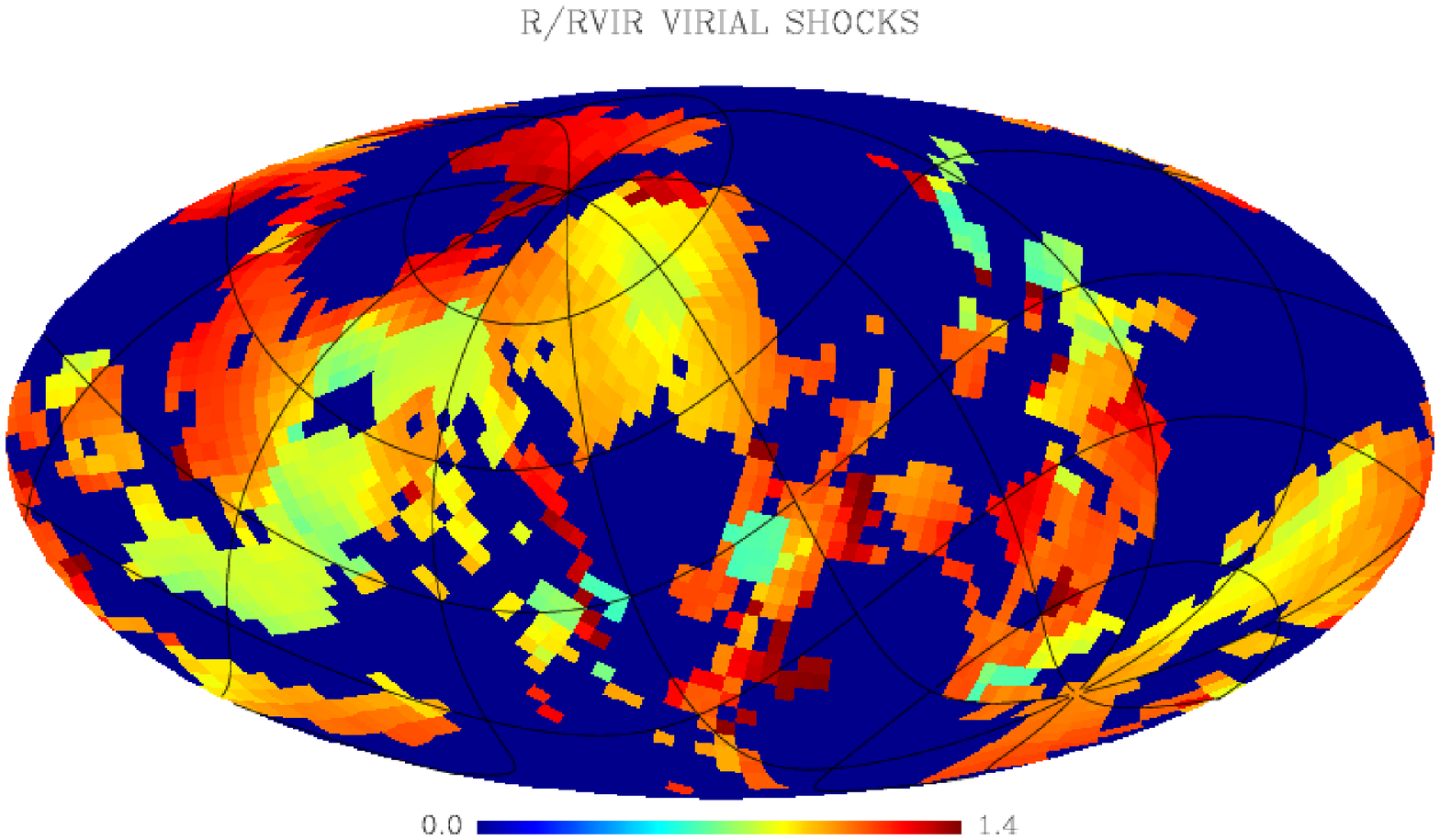}}
     \subfigure[]{
          \includegraphics[width=.9\textwidth]{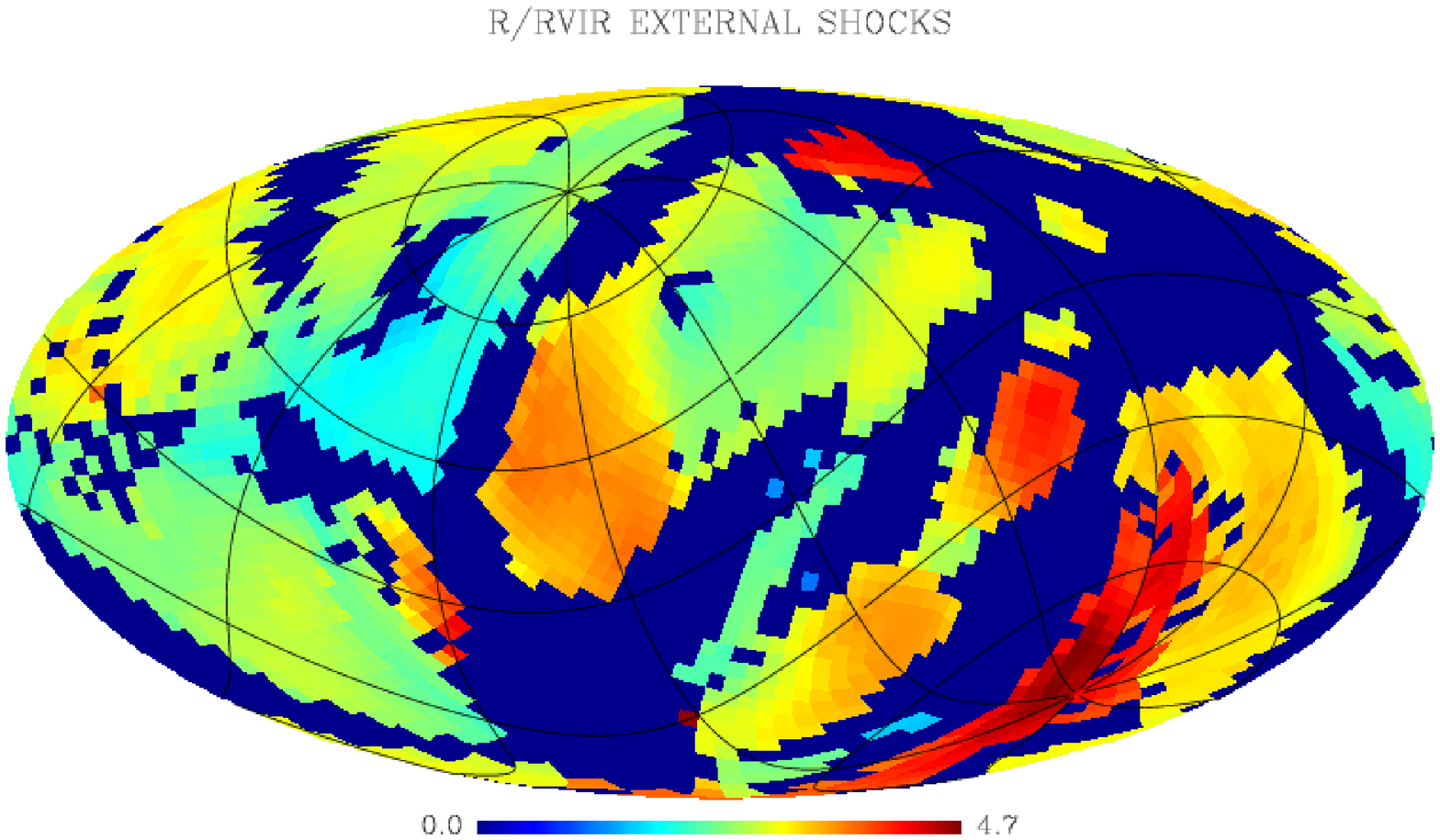}}
\caption{
 Angular distribution of the positions of 
       (a) virial shocks, $R\rmsub{vsh}$ and
       (b) external shocks, $R_x$, 
       around the relaxed AMR cluster AMRCL5 (3072 pixels in \HEALPIX\ projection).
       The color scale shows the radial distance of the shock from the cluster center 
       in units of \RVIR.
\label{F:Fig5}
}
\end{figure}

As an example of how this process works, in Figures~\ref{F:Fig4}c and \ref{F:Fig4}d
we show the histogram of the distributions of shock radii, $R_{x}$, and Mach numbers, $M_x$, 
for external and internal shocks in one relaxed cluster, AMRCL5
(other clusters show similar characteristics). 
Based on the distribution of Mach numbers, shown in Figure~\ref{F:Fig4}d, 
we can identify two populations of shocks: external (solid line) and internal (dashed line) shocks.
In this cluster, the characteristic Mach number for external shocks (solid line)
${\cal M}_x \approx 80$, but the Mach number distribution is broad, 
extending up to very high Mach numbers (10$^4$), 
while the characteristic Mach number for internal shocks (dashed line) is about 3. 
External shocks can be clearly identified by using a lower cut--off Mach number of 10 in this cluster.
The distribution of radial positions of external shocks (including pixels with Mach number
above 10) in this cluster peaks at about ${\cal R}_x = 3$\RVIR, with a width of 1.5\RVIR.
We determine the lower cut--off of external shocks for each cluster by locating the transition 
from internal shocks to external shocks using their Mach number
distribution (typically 10 - 15).
This is clearly an approximation, but it is adequate for our statistical study.

We show the 2D angular distribution of external shocks 
for the relaxed AMR cluster AMRCL5 in Figure~\ref{F:Fig5}b.
The large green and light yellow areas in this figure correspond to 
pixels with $R_x$ around 3 (${\cal R}_x = 3$; see Figure~\ref{F:Fig4}c).
Pixels covered by internal and external shocks are only in rough alignment with one another
due to two reasons: 
i) filaments bend, and become more radial near \RVIR;
ii) the filaments miss the cluster center.

The properties of the external shocks in all AMR clusters are summarized in Table~\ref{T:Table1}, including
their characteristic radial positions, ${\cal R}_x$, and Mach numbers, ${\cal M}_x$, and the
surface area covered, ${\rm A}_x$ (columns $h$, $i$, and $j$, respectively).
This distribution of the radial position of external shocks peaks at about 3\RVIR, 
ranging from 2\RVIR\ to 3.5\RVIR.
The characteristic Mach numbers of external shocks are very high, around 100 or higher, 
due to the low sound speed in the cold, un-shocked gas.
We note, however, that UV background photons, radiation from high--temperature
post-shock gas, and/or other forms of energy injection, which are not included in the adiabatic simulations,
would warm the un-shocked gas to about 10$\,^4$ K, and reduce these Mach numbers.
About half of the surface area around clusters (40\%-60\%) is covered by
these external shocks with very large Mach numbers.

% % % % % % % % % % % % % % % % % % % % % % % % % % % % % % % % % % % % % % % % % % % % % % % % %
\section{Accretion Shocks in SPH Simulations}
\label{S:SPH}

% % % % % % % % % % % % % % % % % % % % % % % % % % % % % % % % % % % % % % % % % % % % % % % % %
\subsection{Locating and Analyzing Accretion Shocks in SPH Clusters}
\label{SS:Shocks}

We use a sample of 10 clusters of galaxies drawn from Lagrangian (SPH) 
simulations (SPHCL1-10, sorted by total virial mass, SPHCL1 being the most massive), 
each located at redshift $z=0$, to analyze accretion shocks.  
The properties of these clusters are summarized in Table~\ref{T:Table2}.
The SPH simulations used preheating (2 keV per particle) to match the observed scalings of 
luminosity and intra-cluster medium mass with temperature 
(Bialek, Evrard \& Mohr, 2001). The underlying cosmology for this 
simulation was a spatially flat, \LCDM\ model with 
($\Omega_{\Lambda}$, $\Omega_m$, $\Omega_b$, $h$, $\sigma_8$) = (0.7, 0.3, 0.03, 0.7, 1.0).
These cosmological parameters are in the range of the parameter values allowed by the \WMAP\ 5 year results 
except that the simulations have a somewhat low $\Omega_b$ and a high $\sigma_8$ \citep{Dunk08}.
The clusters of galaxies were identified in a large, low--resolution simulation
with a box size of 366 Mpc, using dark matter only, and then the regions identified as seeds of 
cluster formation were re-simulated at high resolution, including a full treatment of gas 
dynamics coupled to dark matter by gravity (see Bialek et al. 2001 for details). 
The total virial mass of the clusters in our SPH sample is in the range of 1.1 - 2.1$\times 10^{15}$ \MSUN.
The average resolution (SPH smoothing length) of the high resolution cluster simulations 
at $R/\RVIR$ = 0.1, 1, and 4 were about 40 kpc, 200 kpc, and 1 Mpc, respectively.
As an example, in Figure~\ref{F:Fig1}, we show the spherically averaged resolution 
(SPH smoothing length) in a typical relaxed cluster, SPHCL4 
($\MVIR = 1.4\times 10^{15}~\MSUN$; points with error bars connected with the dashed curve).
We use 768 \HEALPIX\ pixels to represent our SPH cluster data. 
With this choice  the \HEALPIX\ pixel sizes as a function of radius track the resolution of the 
simulations well (within a factor of two; compare the dashed curve with error bars and the straight 
dashed line in Figure~\ref{F:Fig1}).
We apply the same visual method to identify relaxed clusters as in the AMR case above (\S\ref{SS:AMRShocks}).
We find four relaxed clusters in our SPH sample.

In SPH simulations, Lagrangian particles are used as interpolation points to produce 
a continuum representation of the gas.
Given a compact distribution function $W(\vec{x}, h)$, also called a ``smoothing kernel'' 
(usually similar to a Gaussian distribution), a physical quantity, $A$, 
can be expressed at an arbitrary position $\vec{x}$ as
\begin{equation}
 A(\vec{x}) = \sum_{i} A_{i} \frac{m_{i}}{\rho_{i}} W(\vec{x} - \vec{x}_{i} ; h_{i})
,
\end{equation}
where $\rho_{i}$, $\vec{x}_{i}$, $m_i$ and $h_{i}$ are the local mass density, position, 
mass and smoothing length, respectively, associated with particle $i$, 
and the summation is over all particles.

Similarly to our method used to locate shocks in the AMR cluster sample, 
we locate shocks in our SPH clusters using the convergence, $\kappa$.
However, the discreteness of the SPH particles and the unstructured nature of their 
distribution can make regions of strong shear 
(large $\left| \nabla \times \vec{v} \right|$) appear as regions of compression.  
Therefore we used the shear correction factor of Balsara (1995): 
\begin{equation}
f = \frac{\left| \nabla \cdot \vec{v} \right|}{\left| \nabla \cdot 
                     \vec{v} \right| + \left| \nabla \times \vec{v} \right| + \epsilon^{2} c_s / h_{sc}}
\end{equation}
where $\epsilon$ is a small arbitrary constant used to prevent division by zero, 
$c_s$ is the local sound speed, 
and $h_{sc}$ is the scale width of the smoothing kernel.  The factor $f$ is near unity where there is 
no shear, and near zero where the shear is large relative to the compression.  
Thus, shocks should only appear in regions where the corrected convergence, 
$\kappa = - f \, \nabla \cdot \vec{v}$ has large values.
We also looked at the derivative of the internal energy generated by the artificial viscosity (AV),
${D u_{AV}/D t}$.
In order to prevent the unphysical ``post-shock oscillations'' and particle 
inter-penetration, SPH codes typically employ AV to mimic energy dissipation 
and smooth the thermal energy distribution behind a shock front.  
These algorithms are designed to convert kinetic energy into thermal energy only 
around regions of compression, and AV is large only where shocks appear.  
We use the \tmk\ SPH code (Hearn 2002) 
to reconstruct the AV energy derivatives for our SPH cluster data.
The \tmk\ code is an extensible, parallel program written in 
C++, which provides the user with significant flexibility in the simulation 
components used, including artificial viscosity.  
Here, we employ the AV algorithm of Balsara (1995),
and express the contribution of AV to the thermal energy derivative as
\begin{equation}
 { D u_{AV} \over D t} \propto \frac{p}{\rho^2} \left(-\alpha \mu + \beta \mu^2 \right)
,
\end{equation}
where $p$ is the pressure, $\rho$ is the mass density, and $\alpha$ and $\beta$ 
are the viscosity parameters.  
The factor $\mu$ is similar to a velocity divergence; 
the contribution of particle $j$ to the viscosity of particle $i$ is
\begin{equation}
\mu_{i,j} = \left\{ \begin{array}{ll}
                   f_{i,j} \,h_{sc} {\vec{x}_{i,j} \, \cdot \vec{v}_{i,j} \over \left( \left|\vec{x}_{i,j}\right|^{2} 
                                                                    + \epsilon^{2} \right) c_s} & \vec{x}_{i,j} \cdot \vec{v}_{i,j} < 0 \\
                                                                                               0    & \vec{x}_{i,j} \cdot \vec{v}_{i,j} \ge 0 
\end{array} \right.
,
\end{equation}
where $\vec{x}_{i,j} = \vec{x}_{i} - \vec{x}_{j}$, $\vec{v}_{i,j} = \vec{v}_{i} - \vec{v}_{j}$, 
and $f_{i,j}$ is the mean value of the shear correction for particles $i$ and $j$.

\begin{figure}
\centerline{
\plotone{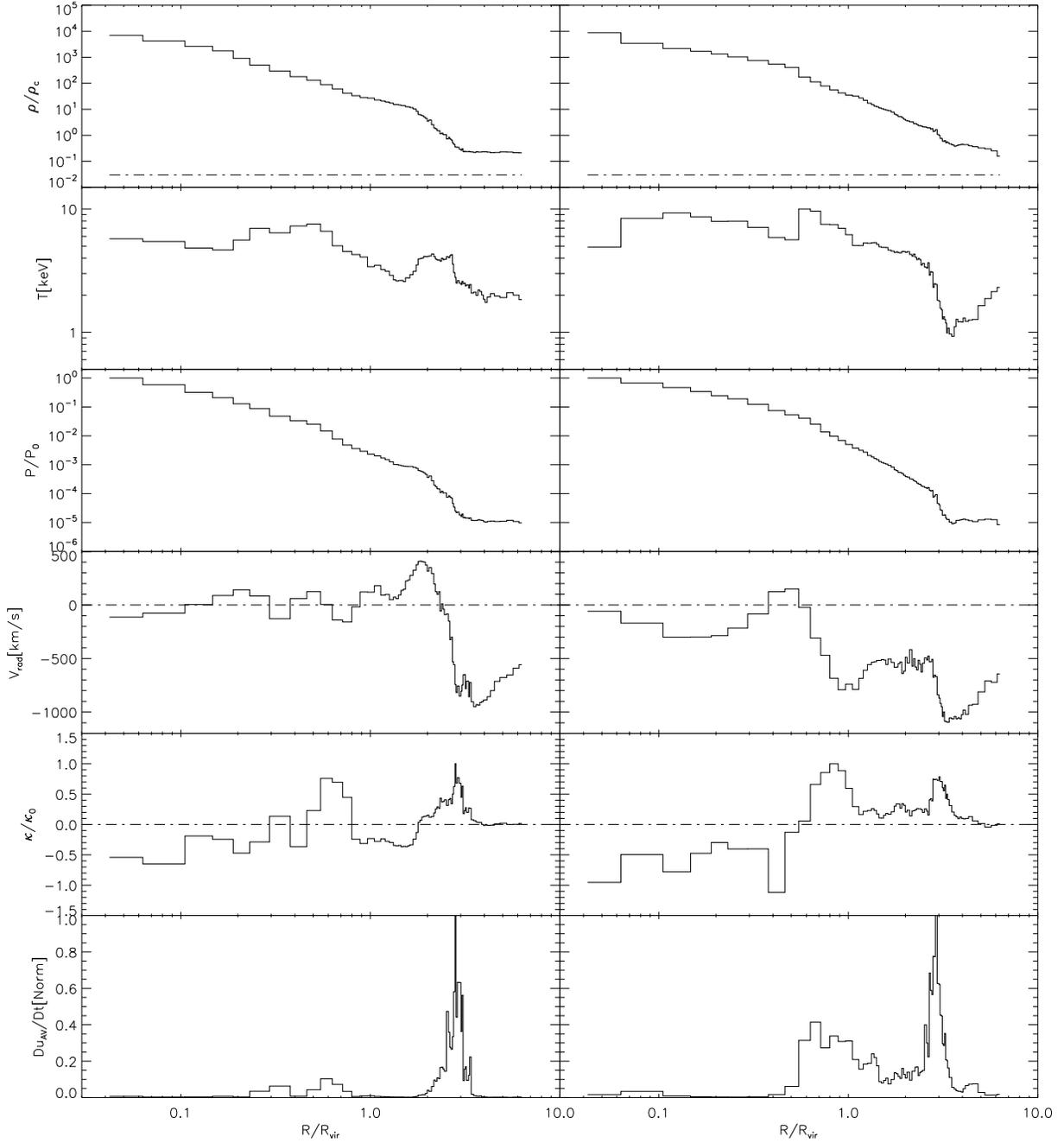}
}
\caption{
 Radial profiles of gas density, temperature, pressure,
 radial velocity, convergence (as in Figure~\ref{F:Fig2}) and ``compression'' 
 (defined as the time--derivative of the internal energy due to artificial
 viscosity, $D u_{AV}/D t$, normalized to its maximum value), from top to bottom.
 Profiles are shown as a function of radius in units of \RVIR\ 
 around the relaxed SPH cluster SPHCL4, in a direction
 avoiding filaments (left panels) and a direction that contains overdense filaments (right panels).
\label{F:Fig6}
}
\end{figure}

We illustrate our results using the relaxed cluster SPHCL4. Other relaxed clusters
show similar characteristics.
This cluster has a total mass of $M = 1.4 \times 10^{15}~\MSUN$, 
and a virial radius of $\RVIR = 2.38$ Mpc.
In Figure~\ref{F:Fig6} we show the distribution of the physical parameters 
toward two fixed pixel directions representing no--filament and filament regions.
In the no--filament regions the largest compression (the maximum of the convergence or
$D u_{AV}/D t$), located at about 3.0\RVIR, is due to the the virial shock. 
However, the radial velocity, \VRAD, becomes zero at around 2.5\RVIR, about 0.5\RVIR\
closer to the cluster center. This offset is due to numerical effects in SPH simulations.
The infalling gas has a very low density, therefore the shock regions in it are smeared out due
to the large smoothing length and artificial viscosity \citep{AbadBN00}.
We find that in the direction of filaments (represented by Pixel B), as we would expect,
the gas plunges deep into the cluster and the radial velocity reaches zero at about 0.6\RVIR.

\begin{figure}
\centerline{
\plotone{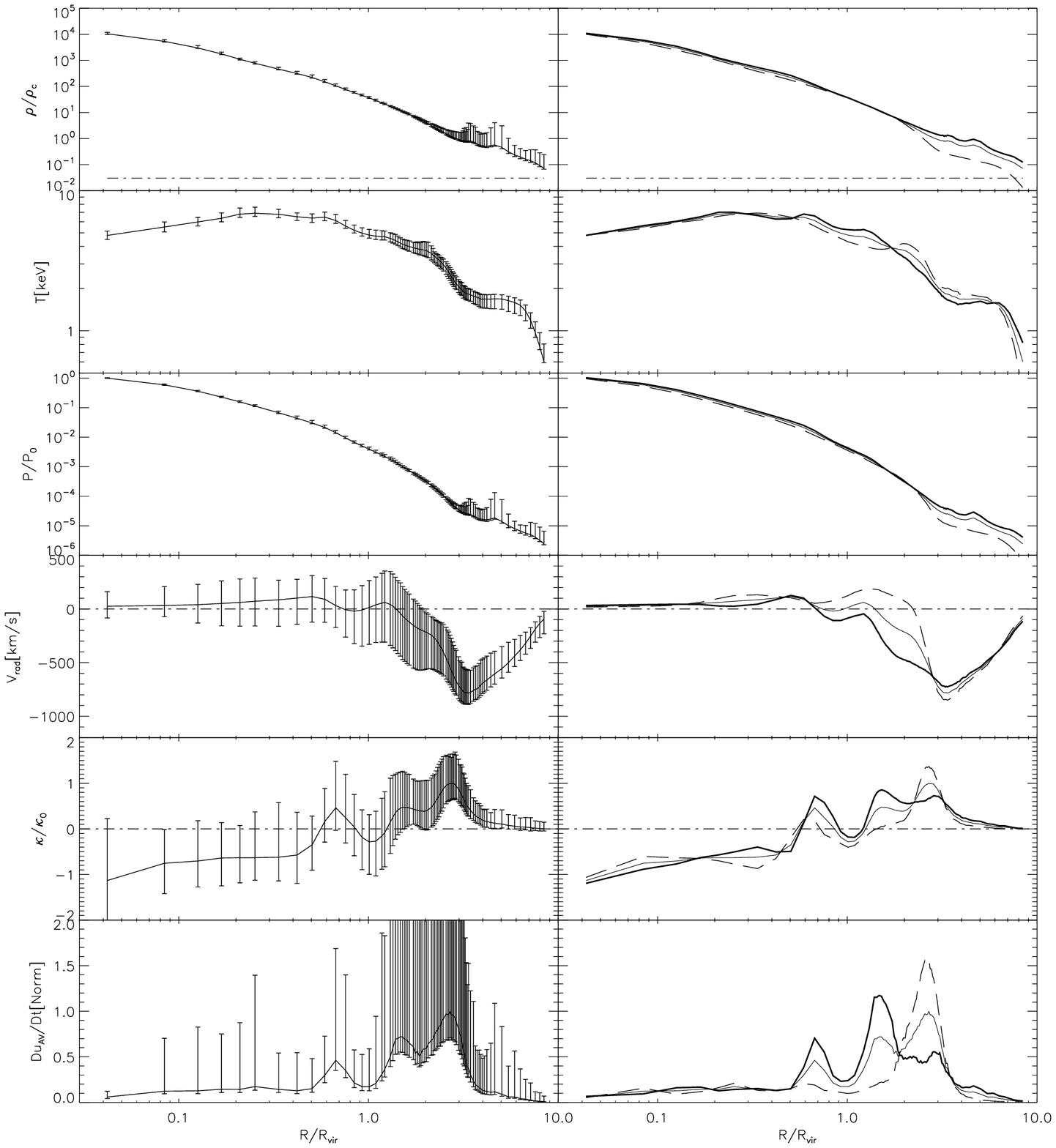}
}
\caption{
Same as Figure~\ref{F:Fig6}, except the left panels show averages over all directions, 
 and the right panels show averages toward directions 
 with filaments (thick solid curves), no filaments  (dashed curves) and, for reference, 
 toward all directions (thin solid lines). 
\label{F:Fig7}
}
\end{figure}

In Figure~\ref{F:Fig7}, we show spherical averages of physical parameters
toward all pixel directions in SPHCL4. 
The left panel shows mean values over all pixel directions with error bars
(the error bars represent the dispersion due to spherical averaging 
calculated as for Figure~\ref{F:Fig3} left panel), 
while the right panel shows averages toward filaments (thick solid curves), no--filaments 
(dashed curves) and, for reference, toward all directions (thin solid curves). 
From top to bottom, we plot profiles of the gas density, temperature, pressure, 
radial velocity, convergence (units are the same as in Figure~\ref{F:Fig2})
and ``compression'' $D u_{AV}/D t$ (in arbitrary units, normalized to its maximum value), 
as a function of distance from the cluster center in units of \RVIR.
The large dispersion on the gas density at large radii of this relaxed cluster
is due to clumps falling into the cluster via filaments, and show that the 
spherically averaged density is not a good approximation far from the cluster center.
The large dispersion between 1\RVIR\ and 3\RVIR\ is due to the fact that
the shock location is a function of direction from cluster center
(most visible in the temperature and the $D u_{AV}/D t$ plots).
The temperature floor at about 2 keV is due to the 2 keV preheating 
assigned to every particle according to the preheating scheme.

From the distribution of the gas radial velocity, \VRAD, (right panel of Figure~\ref{F:Fig7}),
we can see that along the no--filament directions, the gas infall slows down 
at about 3\RVIR, reaching zero velocity at about 2.1\RVIR, whereas in directions toward filaments,
the gas plunges into the cluster to about $0.5~\RVIR$.
The density distribution in the no--filament regions (dashed line) drops below the 
average density (solid line) at around 2.1\RVIR.
Based on our simulations, this should be the position of the virial shock in this cluster.
The position of the maximum of the convergence and $D u_{AV}/D t$ 
(last two figures on the right panel) is at about 2.7\RVIR. 
This shows that the maximum compression happens at about 0.6\RVIR\ farther
from the cluster center than where the virial shock position should be
based on the density and velocity distributions.
This is due to the large smoothing length (about 1 Mpc at this distance) and the artificial viscosity.

Since the density is very low in the pre-shock regions, the large smoothing length and the artificial 
viscosity make it difficult to estimate the pre-shock values of density, temperature, radial velocity,
at the shock reliably, therefore we omit the analysis of Mach numbers for our SPH cluster sample.

% % % % % % % % % % % % % % % % % % % % % % % % % % % % % % % % % % % % % % % % % % % % % % % % %
\subsection{Results: SPH Clusters}
\label{SS:SPHResults}

% % % % % % % % % % % % % % % % % % % % % % % % % % % % % % % % % % % % % % % % % % % % % % % % %
\subsubsection{Virial Shocks}
\label{SSS:VShocksSPH}

\begin{figure}
\centerline{
\plotone{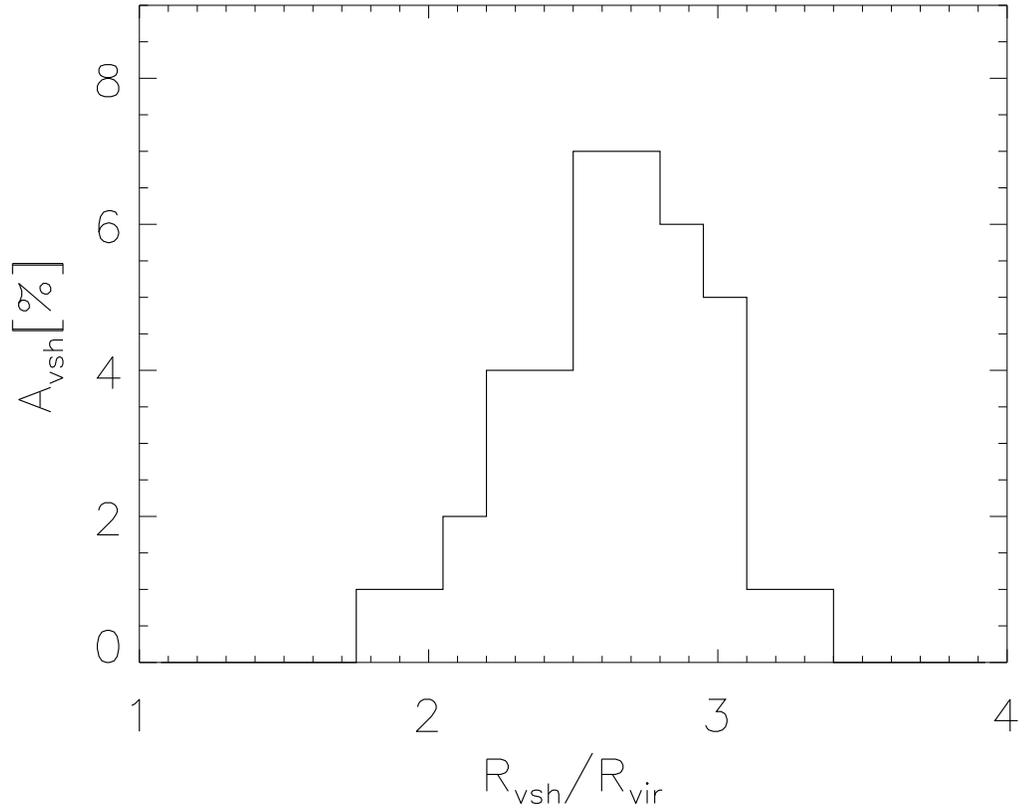}
}
\caption{
 Histogram of the radial location of virial shocks, $R\rmsub{vsh}$, 
 expressed as the percentage of the total surface area occupied by
 shocks at this distance from the center of the relaxed cluster SPHCL4.
\label{F:Fig8}
}
\end{figure}

\begin{figure}
\centerline{
   \includegraphics[width=.9\textwidth]{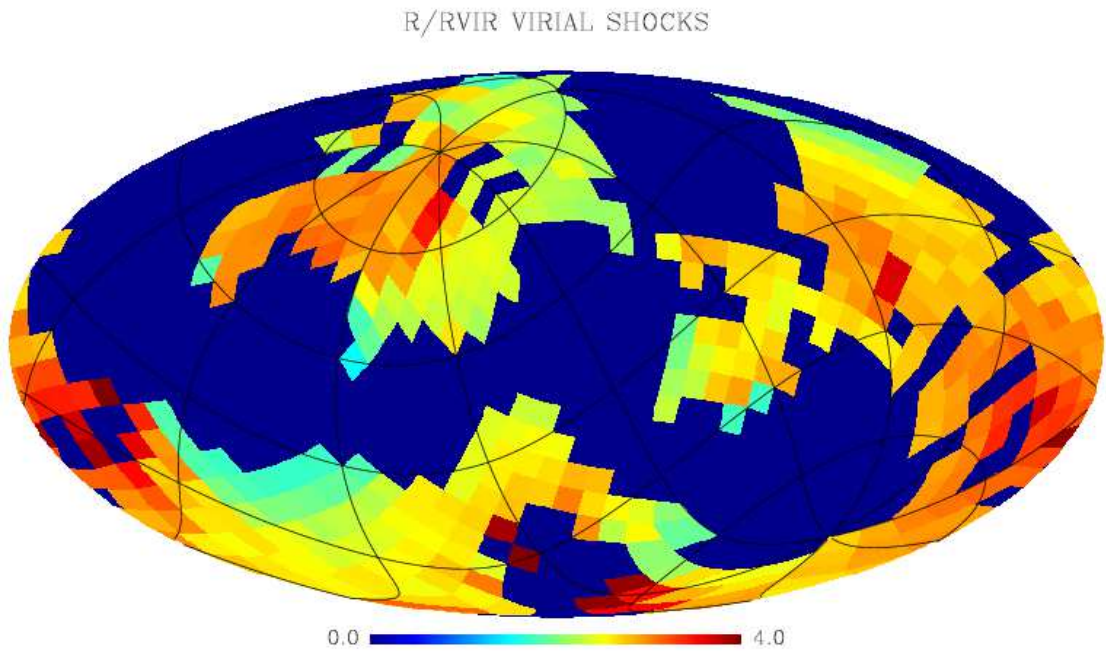}
}
\caption{ 
  Angular distribution of the radial positions of virial shocks, $R\rmsub{vsh}$, around the
  relaxed SPH cluster SPHCL4 (768 pixels in \HEALPIX\ projection),
  with the color scale in units of \RVIR\ as in Figure~\ref{F:Fig5}.
\label{F:Fig9}
}
\end{figure}

The method we apply to locate virial shocks in our SPH sample is similar to that 
applied in the AMR sample. 
We use the convergence maximum and the radial velocity to identify the position of
the virial shocks to be consistent with our analysis of the AMR simulations.
We illustrate our SPH results on one relaxed cluster, SPHCL4, but other relaxed 
clusters show similar characteristics.
In Figure~\ref{F:Fig8}, we show the radial distribution of virial shocks toward different pixels 
expressed as a percentage of surface area covered by shocks 
(number of pixels with shocks over total number of pixels).
The distribution peaks at about 2.7\RVIR\ with a width of 1\RVIR.
The 2D angular distribution of the virial shocks in SPHCL4 is shown in
Figure~\ref{F:Fig9}.

The characteristic positions, ${\cal R}\rmsub{vsh}$,
and the fractional areas covered by virial shocks, ${\rm A}\rmsub{vsh}$, for clusters in our SPH sample are listed
in Table 2 (columns $e$ and $f$).
The positions of the virial shocks in most preheated SPH clusters fall between 
2\RVIR\ and  3\RVIR.
In relaxed clusters, the virial shocks cover a fraction 36--56\% of the total $4\pi$ solid angle.

% % % % % % % % % % % % % % % % % % % % % % % % % % % % % % % % % % % % % % % % % % % % % % % % %
\subsubsection{External Shocks}
\label{SSS:XShocksSPH}

Unlike in the AMR case, in the no--filament directions, the SPH clusters always show 
only a single accretion shock in each line of sight.  Since there is no sign of a shock farther out, 
and the kinetic energy from radial infall is being converted to thermal energy by these shocks, 
we refer to these as virial shocks.

However, at about 3\RVIR, which is where we would expect to find the
external shocks in our AMR simulations, the pre-shock gas in the SPH simulations 
experiences strong numerical smoothing (as already emphasized in \citealt{AbadBN00}).
The smoothing length at these large distances is greater than 1 Mpc, thus even if 
external shocks were formed at these locations, the SPH simulations could not resolve it. 
Therefore we do not analyze external shocks in our SPH cluster sample.

% % % % % % % % % % % % % % % % % % % % % % % % % % % % % % % % % % % % % % % % % % % % % % % % %
\section{Discussion}
\label{S:Discussion}

% % % % % % % % % % % % % % % % % % % % % % % % % % % % % % % % % % % % % % % % % % % % % % % % %
\subsection{Observing Virial Shocks Using Their SZ Signature}
\label{SS:ObserveSZ}

The primary motivation in this paper for studying accretion shocks -- besides characterizing their basic properties -- 
is to assess whether they could be detectable by \ALMA\ through the SZ effect. 
As stated in the Introduction,
the advantage of SZ observations is that the SZ surface brightness is proportional to the gas 
density (rather than its square, such as the X--ray flux), and therefore
drops considerably less rapidly with radius.
Using spherically symmetric models for the cluster gas and temperature profiles, 
\cite{Kocset05} calculated the signal-to-noise ratio ($S/N$) for detecting
shocks near the virial radius of clusters, using the high spatial resolution ($2\,\arcsec$), high sensitivity 
($\approx 10\mu$K, with a day of integration) next generation interferometer, \ALMA.
They defined the $S/N$ ratio as the difference in the predicted surface brightness near the 
location of the shock, in the presence and absence of a strong discontinuity in gas pressure, divided by
the instrumental noise. The Kocsis et al. study was ``semi--analytical'',
based on a spherically symmetric, self-similar, polytropic intra-cluster gas model, in hydrostatic
equilibrium in a Navarro, Frenk \& White (1997; NFW) dark matter halo.
Different virial shock positions were assumed between 1 and 2\RVIR, where \RVIR\ is the 
fiducial virial radius.
Non-spherical accretion and projection effects were taken into account 
by assuming that the shock fronts are spread over a finite radial distance, 
with a width of $0 < \Delta R < \RVIR$.

For the sake of completeness, we briefly recapitulate the pertinent aspects of 
the \citet{Kocset05} model, including their calculation of the SZ surface brightness.
The interested reader should consult their work for more details.
The SZ surface brightness at the impact parameter $y$ relative to the center
of a cluster with a given mass \MVIR\ and redshift $z$ is expressed as 
\begin{equation}
   \Delta T(y) =  \Delta T_s \, Y(y),
\end{equation}
where $y$ is the dimensionless distance from the cluster's center at
the closest approach for a given sightline (in units of the virial radius),
and $\Delta T_s$ is a constant normalization,
\begin{equation}
   \Delta T_s = p(x_{\nu}) \, { \sigma_T \, T\rmsub{CMB} \over m_e \, c^2} \, { G \MVIR \over 3 \, c\rmsub{NFW} } \,
                        \eta \, \rho_g(0).
\end{equation}
Here, the frequency dependence is given by $p(x_{\nu}) = x_{\nu} \coth x_{\nu}/2 - 4$ 
(e.g. Birkinshaw 1999), where the dimensionless frequency is $x_{\nu} = h_P \nu / k_B T\rmsub{CMB}$ 
($h_P$ and $k_B$ are the Planck and Boltzmann constants; relativistic corrections are 
ignored since they are negligible for \ALMA\ observations at $\nu = 100$ GHz of clusters 
with $T_g \simless 10$ keV); $T\rmsub{CMB}$ is the temperature of the CMB today; 
$c\rmsub{NFW}$ is the concentration parameter for the NFW halo profile; 
$\rho_g(0)$ is a normalization constant for the gas density, determined by the universal baryon 
fraction; and $\eta$ is a normalization constant for the mass--temperature relation.
Assuming a polytropic equation of state $P\rmsub{gas}\propto\rho^\gamma$,
and also that the gas density profile tracks that of the dark matter
at large radii, determines the polytropic index, $\gamma$, and the
normalization, $\eta$ (Komatsu \& Seljak 2001).

The shape of the surface brightness profile is contained in
\begin{equation}
    Y(y) =  2 \, \int_0^{\ell\rmsub{cut}(y)} W(x) \, y_g(x)^\gamma d\ell,
\end{equation}
where the integral is performed along the line of sight $\ell$; $x =
\sqrt{y^2 + \ell^2}$ is the dimensionless radial coordinate (we use
\RVIR\ as a unit for distance, thus $x = r / \RVIR$); $\ell\rmsub{cut}(y) =
\sqrt{ (\XMAX + D)^2 - \ell^2}$, where \XMAX\ is the assumed radial
position of the virial shock with a width of $2 D$, $y_g(x)$ is the
solution of the equation of hydrostatic equilibrium with a polytropic
gas model (Komatsu \& Seljak 2001), and $W(x)$ is a linear
cut--off function that allows for the edge of the cluster to be ``blurred'',
mimicking shocks that are spread over a finite distance:

\begin{equation}
     W(x) = \left\{ \begin{array}{ll}
                  1                                          &   x < \XMAX - D                            \\
                  { \XMAX + D - x \over 2 D} &   \XMAX - D \le x \le \XMAX + D \\
                  0                                          &   x > \XMAX + D 
\end{array} \right.
.
\end{equation}
\citet{Kocset05} used the above model to compute the difference $\delta
\Delta T$ in surface brightness profiles for pairs of clusters, one
with no edge ($\XMAX\rightarrow\infty$), and another cluster that
was identical, except that it has an edge near the virial radius
($\XMAX\approx 1$).  By comparing $\delta \Delta T$ to the noise
expected from \ALMA, they showed that, in theory, virial shocks
should be detectable with high significance.

We find that the positions of virial shocks in our simulated AMR
clusters fall in the distance interval of 0.9-1.4\RVIR\ (see Table~\ref{T:Table1}).
Here we focus on these virial shocks, rather than any external shocks,
because the SZ surface brightness at $\approx 3 \RVIR$ is likely to be
too low to yield a detection of any external shocks.
From Figure~\ref{F:Fig4} we can see that over different
directions, the radial position of the virial shocks varies by about
$\pm 0.5\RVIR$.  The effective width $D$ of these shocks (which do not
include the directions along filaments) will be narrower, since the 2D
projected SZ surface brightness in a given direction on the sky
depends only on a small fraction of the entire 4$\pi$ steradian solid
angle around the cluster.  Virial shocks cover about 40\% of the total
surface area of clusters (see Table~\ref{T:Table1}), rather than the 100\% coverage
assumed by \cite{Kocset05}.

\begin{figure}
     \centering
     \subfigure[]{
          \includegraphics[width=.45\textwidth]{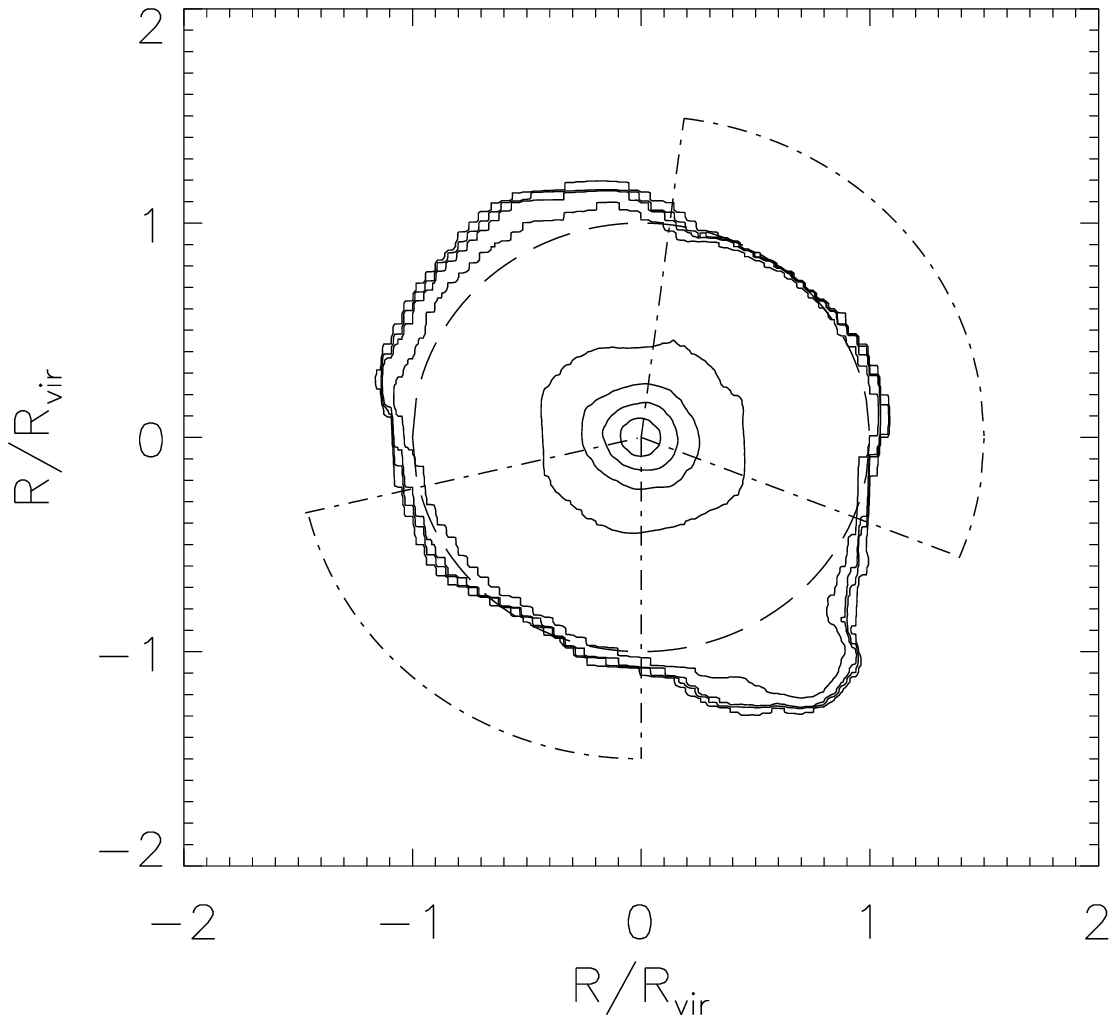}}
    \subfigure[]{
          \includegraphics[width=.45\textwidth]{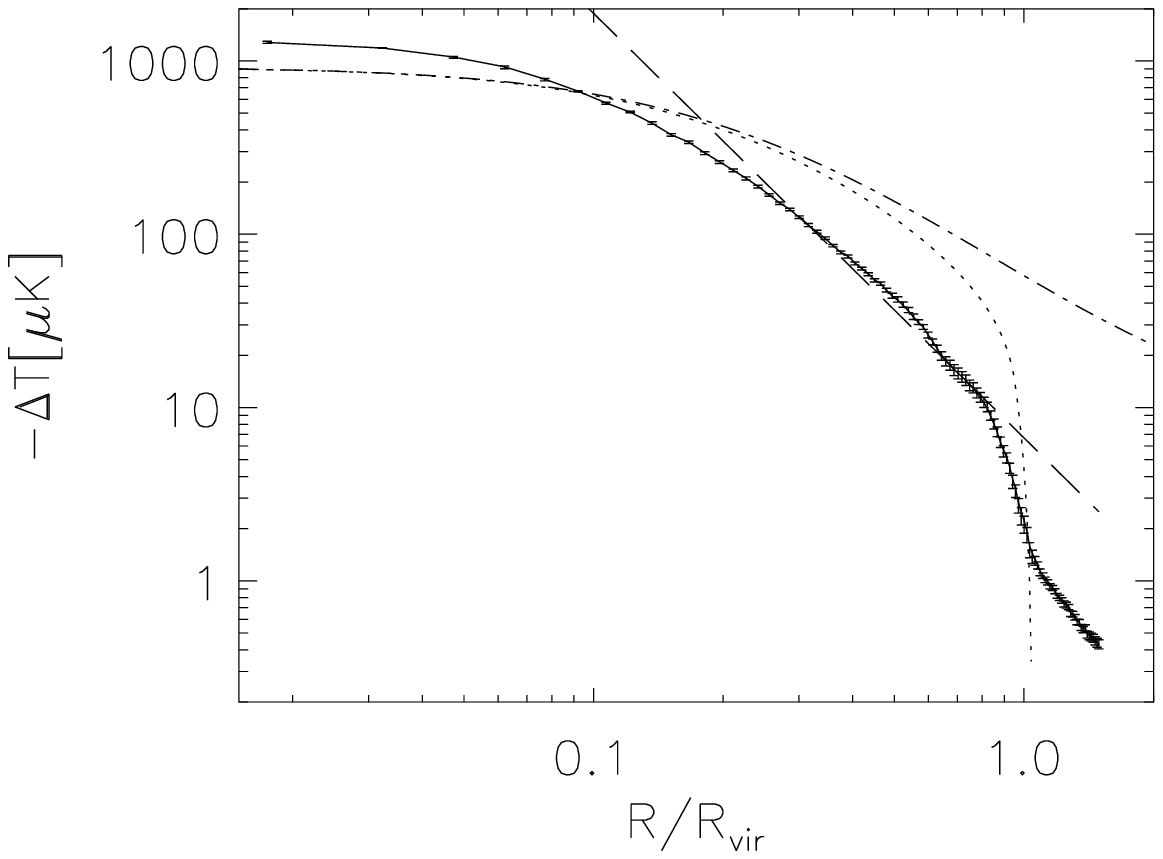}}
     \subfigure[]{
          \includegraphics[width=.45\textwidth]{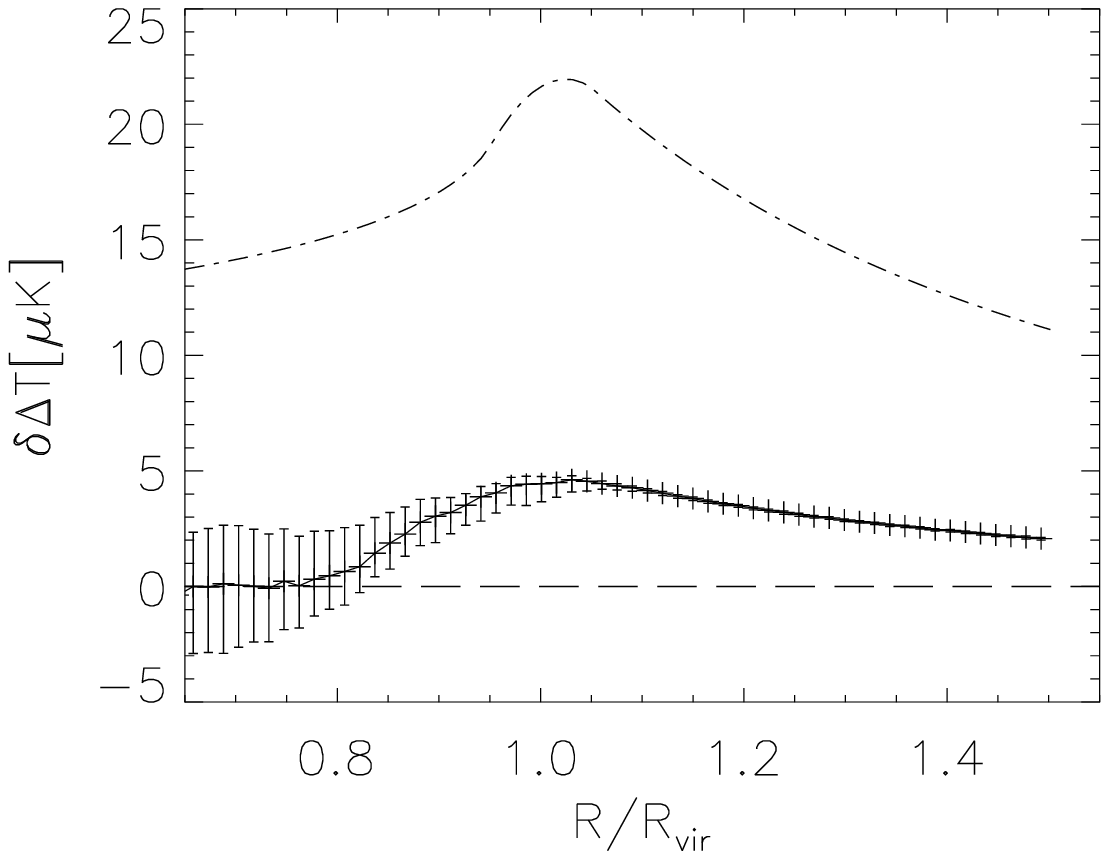}}
\caption{
   SZ signature of virial shocks in the relaxed massive cluster AMRCL5, computed
   at the frequency of 100 GHz.  The cluster is viewed from a direction that
   minimizes contamination from filaments.  Panel (a) shows 2D contours of the
   SZ decrement (solid curves). 
   The contour levels correspond to 
   $-\Delta T$ = (1.5, 1.7, 1.9, 2.8, 60, 190, 380, 760) $\mu$K.
   The wedge--shaped regions enclosed by the dot--dashed curves are free of filaments.
   Panel (b) shows radial SZ profiles, averaged azimuthally in the no--filament regions.
   The solid curve with error bars shows the simulated profile; the long--dashed
   line is a power--law fit to this profile near \RVIR. The dotted and dot--dashed
   curves show predictions of a semi--analytical toy model with and without a virial
   shock, respectively. Panel (c) shows the difference between the SZ profiles 
   with and without a virial shock.  The solid curve with error bars is the difference
   between the simulated profile and the power--law fit; the dot--dashed curve is the
   difference between the pair of toy--models in panel (b).
\label{F:Fig10}
}
\end{figure}

\begin{figure}
     \centering
     \subfigure[]{
          \includegraphics[width=.45\textwidth]{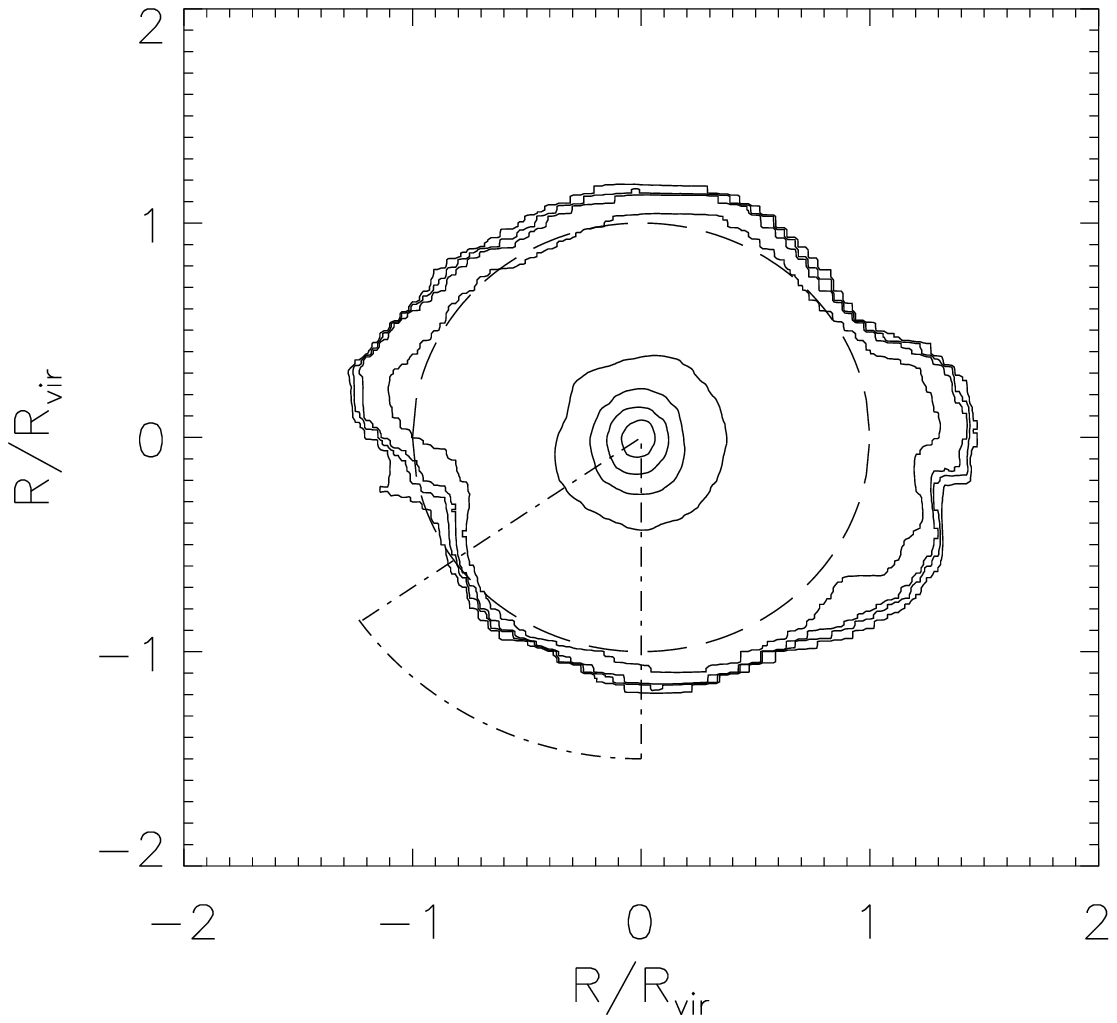}}
    \subfigure[]{
          \includegraphics[width=.45\textwidth]{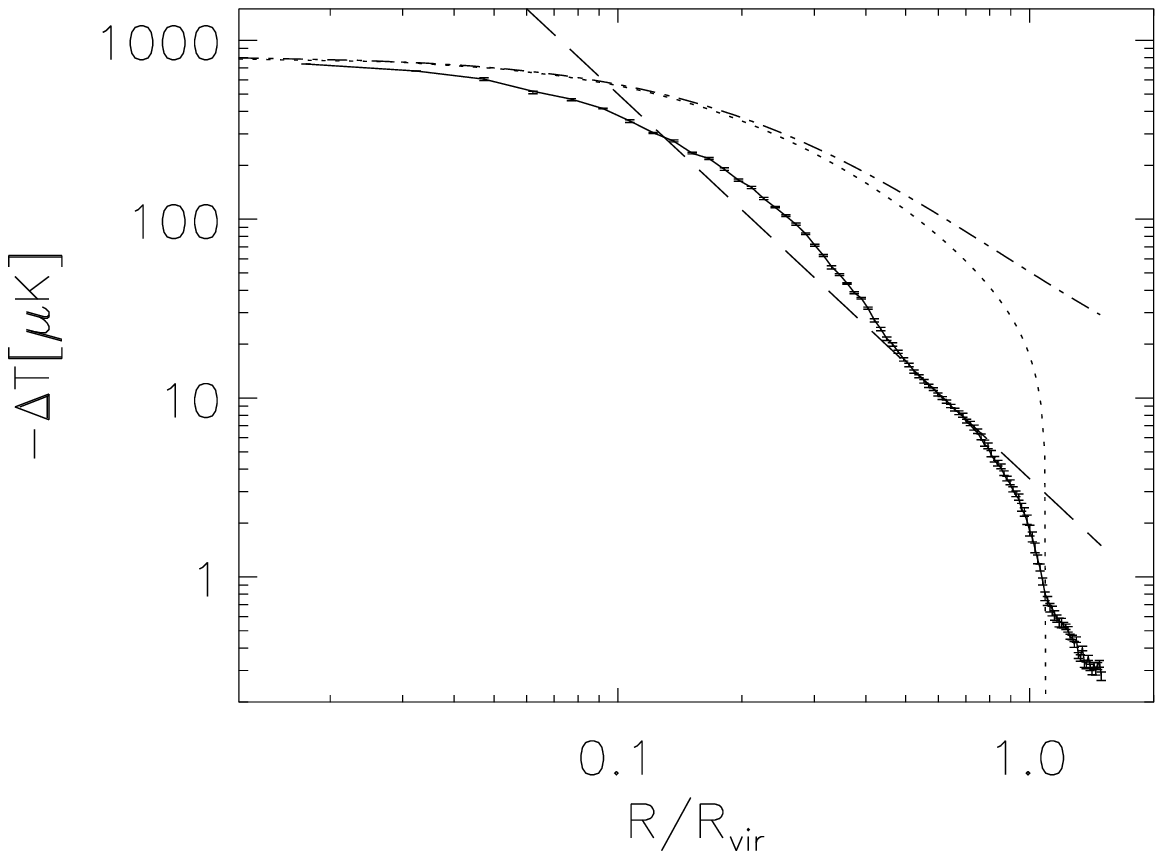}}
     \subfigure[]{
          \includegraphics[width=.45\textwidth]{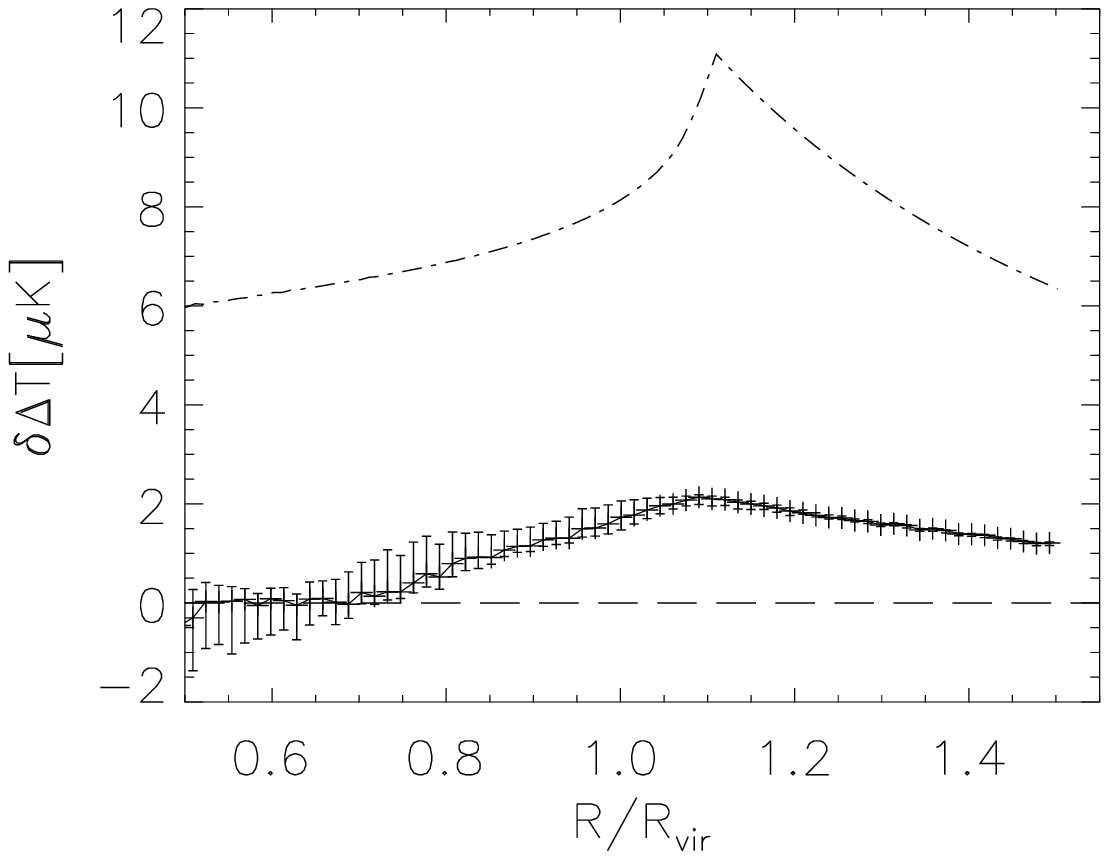}}
\caption{
 Same as Figure~\ref{F:Fig10}, except for the massive cluster AMRCL7 that is undergoing a merger,
   viewed from a direction that maximizes contamination from filaments.  
   The contour levels correspond to $-\Delta T$ = (0.9, 1.0, 1.1, 1.6, 32, 110, 220, 430) $\mu$K.
}
\label{F:Fig11} % Figure~\ref{F:Fig11}
\end{figure}

Most importantly, the 2D projection of the surface area covered by
virial shocks can change substantially, depending on the viewing angle
relative to the active filament plane of the cluster.  Here ``active
filament plane'' refers to the plane crossing the cluster center that
contains most of the filaments around the cluster. The orientation of
this plane can be determined unambiguously by visual inspection in almost 
all clusters in our sample, and we expect that this plane can indeed be identified 
in most relaxed massive clusters (but our sample of relaxed clusters is not large 
enough to verify this).
For example, 50\% of the 2$\pi$ azimuth angle measured from the
cluster center is covered by virial shocks in the XZ-projection of
AMRCL5 (Figure~\ref{F:Fig10}a; area within dot-dashed lines) whereas this
fraction is only 10\% in the YZ-projection of AMRCL7 (Figure~\ref{F:Fig11}a; area
within dot-dashed lines).  The regions contaminated by filaments in
these 2D images were identified via visual inspection.  Since the
majority of filaments around rich clusters of galaxies are located in
the active filament plane of the cluster, the contamination by
filaments can be minimized quite effectively by choosing clusters that
we happen to observe ``edge on'', i.e. we are located in the active
filament plane. Filaments might be identifiable using galaxy redshift
information (see Pimbblet, Drinkwater \& Hawkrigg 2004), which would help in picking the
most promising clusters for observations with \ALMA.

We next use two massive clusters, AMRCL5 and AMRCL7, to make 
explicit comparisons to the spherical model predictions in \citet{Kocset05},
and to illustrate some issues related to observing virial shocks based
on their SZ signature (Figures~\ref{F:Fig10} and \ref{F:Fig11}).  
Relaxed cluster AMRCL5 has a total mass of $1.2 \times 10^{15}~\MSUN$, 
and AMRCL7, a merging cluster, has a mass of $1.1
\times 10^{15}~\MSUN$ (see Table~\ref{T:Table1}).

We first calculate the 2D SZ surface brightness distribution from our
simulation data, integrating along the line of sight ($\ell$) over the
extent of the cluster along the line of sight (from $\ell_1$ to
$\ell_2$) using

\begin{equation}
           \Delta T(x,y) =  p(x_{\nu}) \, T\rmsub{CMB} \,{k_B \sigma_T \over m_e c^2}
                                                     \int_{\ell_1}^{\ell_2} n_e(x,y,\ell) \, T_e(x,y,\ell) \, d\ell
.
\end{equation}
Here $(x,y)$ are spatial coordinates in the plane of the sky,
i.e. perpendicular to the line--of--sight coordinate $\ell$; $n_e =
f_g \rho_g / \mu_e m_p$ is the electron density, where $\mu_e$ is the
mean molecular weight per electron and $m_p$ is the proton mass;
$\rho_g$ is the gas density; $f_g$ is the gas mass fraction (we adopt
$f_g=0.9$), and we use the standard assumption that the electron
temperature equals the gas temperature, $T_e = T_g$ (similar
expressions were used for the XZ and YZ projections).  In practice, we
pixelize $x,y$ and $\ell$, and approximate the integral with a sum
over the line of sight from $\ell_1 = -10$ Mpc to $\ell_2 = 10$ Mpc.
\cite{Kocset05} found that the $S/N$ ratio is the lowest for clusters
at a redshift of $z=0.3$, thus we generated images of clusters
assuming they are located at this redshift. This choice makes our
$S/N$ ratio estimates conservative.

We first examine the XZ projection of AMRCL5, which, as mentioned
above, is close to an edge on projection, and therefore has the least
amount of contamination from filaments (Figure~\ref{F:Fig10}).  In this
projection, the line of filaments is stretching from the upper left
corner to the lower right corner of the image (as revealed by the
shape of the contours in Figure~\ref{F:Fig10}a).  We find that other views of
even relaxed clusters are often much more contaminated by filaments.
The azimuthally averaged SZ surface brightness profile, excluding the
angles contaminated by filaments (i.e. the area within the dot-dashed
lines, containing $\approx$50\% of the 2$\pi$ azimuth angles from the
cluster center) is shown in Figure~\ref{F:Fig10}a. In this figure,
the error bars represent the dispersion (standard deviation) due to azimuthal averaging 
at each radius (instrument noise and contamination from other astrophysical effects are 
not included).

In Figure~\ref{F:Fig10}b, we see that the virial shock appears to
leave a relatively sharp feature in the average SZ profile, i.e. a
sharp drop near \RVIR\, with a width of $\approx$ 0.1\RVIR.  In
addition, there is a distinct break in the slope of the surface
brightness profile at $\approx$0.6\RVIR. This break is not caused by
the change in the slope of the NFW density profile, since it is not
present in the spherical models.  We suspect it is caused by the
breaking of the self-- similarity of the pressure profiles due to the
finite entropy of the cluster gas.  As a consequence, the spherical
models discussed above, shown by the dotted and dot--dashed curves in
Figure~\ref{F:Fig10}b (with and without an edge, respectively), do
not fit the simulated SZ profiles over the whole region inside the
virial shocks.

To quantify the significance at which the presence of the virial shock
can be detected, we fit a power--law model to the SZ surface
brightness profile near the virial shock, as might be done to an
actual observation (dashed curve).  The difference between the
simulated SZ profile and the power--law model ($\Delta T\rmsub{pow} \propto
r^{-\alpha}$, $\alpha$ = 2.45) with no cut--off, $\delta \Delta T =
\Delta T_{cl} - \Delta T\rmsub{pow}$, is shown in
Figure~\ref{F:Fig10}c.  We also show the difference (dot--dashed
curve) between the SZ profiles in the Kocsis et al. spherical models for a
cluster with the same mass, located at a redshift of $z=0.3$, with and
without an edge (dotted and dot-dashed curves in
Fig.~\ref{F:Fig10}b; cf. Figure 2 of Kocsis et al. 2005).  For the
cluster with an edge, we chose the position and width of the virial
shock to be similar to those of our simulated cluster ($\XMAX =
1\RVIR$ and $D = 0.1\RVIR$).

The central SZ decrement of our simulated cluster is $\Delta T_0 =
-1360 \mu$K.  The difference in the SZ decrement of our simulated
cluster, compared to the power--law extrapolation, is about $\delta
\Delta T\approx 4\mu$K within a 0.2\RVIR\ wide annulus around the
1\RVIR\ (evaluated in the no--filament regions).  Assuming an angular
resolution of $2\,\arcsec$, and taking into account the 50\% coverage,
this corresponds to $3.4 \times 10^4$ independent pixels with a virial
radius of $490\,\arcsec$, at z=0.3.  Following \cite{Kocset05}, we
assume 20 hour on-source integration time, and estimate the $S/N$
ratio as $S/N \approx N\rmsub{pix}^{1/2} (S/N)_1$, where $N\rmsub{pix}$ is the
number of independent pixels and $(S/N)_1$ is the signal to noise for
one pixel, and we obtain a high $S/N$ ratio of about 70.  The Kocsis
et al. model has a somewhat lower central decrement of -921$\mu$K, but
it has a significantly flatter SZ profile, and it predicts a factor of
$\approx 5$ higher $S/N$ than our analysis based on the XZ projection
of AMRCL5.

For comparison, we next estimate the $S/N$ ratio for detecting virial
shocks in a cluster which is {\em not} optimally selected to be edge
on.  This case is illustrated in Figure~11a, showing the YZ projection
of the 2D SZ surface brightness map of AMRCL7. Again, the dashed
circle represents the projected \RVIR, and the area enclosed by the
dot-dashed wedge show the region where virial shocks can be seen (no
projected filament contamination).  As we expect, only a small
fraction, in this case 10\% of the 2$\pi$ azimuth around the cluster
center, is not contaminated by filaments.  The azimuthally averaged SZ
profile of this cluster in this no--filament region is shown in
Figure~\ref{F:Fig11}b.  The central SZ decrement for this cluster
in this projection is $\Delta T_0 = -736 \mu$K.  This cluster also
shows a sharp drop in the surface brightness profile near \RVIR\, and
also a discernible break in the slope at around 0.4\RVIR, somewhat
closer to the cluster center than the break in AMRCL5.  The difference
between the simulated surface brightness and the extrapolation of a
power--law fit (with slope $\alpha$ = 2.15), $\delta \Delta T$, is
shown in Figure~\ref{F:Fig11}c.  This shows $\approx$2$\,\mu$K
difference in a 0.2 \RVIR\ wide annulus.  Again, assuming an angular
resolution of $2\,\arcsec$, and taking into account the 10\% coverage,
this corresponds to $6.2 \times 10^3$ independent pixels with a virial
radius of $470\,\arcsec$, at a redshift of $z=0.3$.  Using a simple
estimate as above, we obtain $S/N\approx 16$ for the detection of the
virial shock in this cluster.  The Kocsis et al. model with the same
mass and $\XMAX = 1.1\RVIR$, $D = 0.05\RVIR$ located at a redshift
of $z=0.3$ has a central SZ decrement of -809$\mu$K, and yields a
$S/N$ ratio about 10 times higher than our model.

Based on our analysis of the SZ profiles of simulated clusters, we
have found that the $S/N$ ratios for detecting virial shocks in
different clusters are quite significant, $S/N \approx 16-70$, but
reduced by a factor of 5-10 relative to the predictions of the
spherically symmetric, self-similar models used by \cite{Kocset05}.  
There are two reasons for this reduction: (i) the simulations show a
generally steeper SZ profile, especially at radii beyond 0.5 \RVIR;
and (ii) only a fraction (10-50\% in our two cases) of the
2$\pi$azimuthal angle around the cluster's perimeter is covered by
virial shocks near \RVIR\ in our simulated clusters.

The second of these effects is simply caused by the smearing of the
shocks in the 2D projection of the three--dimensional, non-spherical
features, over a significant fraction of the cluster's perimeter.  The
first effect is attributable to two factors. First, we find that the
concentration parameter for our massive relaxed clusters is about
$c\rmsub{NFW}=5.5$, much larger than the value $c\rmsub{NFW}=2.4$
adopted by Kocsis et al. Second, the drop in the surface density
profile might be related to the excess entropy in the core of the
cluster. (The slope of the SZ profile could thus plausibly provide
information on the entropy structure of clusters; a break in its slope
may be verified by \ALMA\ relatively easily, since the SZ decrement is
much larger at around 0.5 \RVIR\ than at the virial radius.)  Finally,
we note that the toy model used in Kocsis et al. adopted a somewhat
different outer boundary condition than was used in the original work
of Komatsu \& Seljak (2001), on which the Kocsis et al. toy models are
based. In particular, while Komatsu \& Seljak require the {\it slope}
of the gas profile to match that of the DM halo at the virial radius,
Kocsis et al. require the baryon/DM mass ratio interior to a large
radius (=200 $R_{\rm vir}$) to equal the universal mean
$\Omega_b/\Omega_{\rm DM}$.  As a result, the SZ profile in the Kocsis
et al. toy model is significantly flatter near the virial radius than
that of the Komatsu \& Seljak toy model (e.g. the drop in the surface
brightness profile from the cluster core to the virial radius,
predicted in Figure 1 in Kocsis et al., is a factor of $\sim 50$,
which is about a factor of $\sim 2$ smaller than the factor of $\sim
100$ drop predicted for the same cluster parameters in the upper left
panel in Figure 15 in Komatsu \& Seljak).

Our simplified treatment suggests that even if we do not select
clusters optimally (to minimize contamination by filaments), \ALMA\
may still be able to detect virial shocks with
mosaic observations covering
a significant portion of the annulus near \RVIR\ for a suitably bright cluster.
If in real clusters the surface brightness drops below 10$\mu$K around
the virial radius and $\delta \Delta T$ is only about 2-3$\,\mu$K,
then the detection of virial shocks will become much more difficult,
due to contamination from secondary effects in the CMB at the $\mu$K
level from unrelated background or foreground galaxy clusters, dust
emission, and point source contamination.  Additional secondary
effects which can be important at around \RVIR\ may be generated by
clusters of galaxies via gravitational lensing, and by the thermal SZ
effect from unresolved clusters.  A bipolar pattern with an amplitude
of a few $\mu$K is produced by clusters when they lens CMB
fluctuations with a large--scale gradient \citep{SeljZald00}. A
similar pattern is generated by the so--called moving cluster effect
\citep{BirkGull83}, but the amplitude of this effect at the virial
radius is most likely below $1\mu$K \citep{MolnBirk00}.  The
theoretical sensitivity of interferometers is also reduced for
large--scale smooth surface density distributions, therefore we may
have overestimated the $S/N$ ratio.  On the other hand, using
nonlinear de-convolution, much larger scale structures can be
recovered from mosaic observations than an analytical treatment would
suggest \citep{Helfet02}.  
 A redshift of 0.3 was used to obtain a conservative estimate of the $S/N$ ratio;
 we expect a significant increase in $S/N$ if the redshift of the clusters were selected optimally.
An improved estimate for the significance
of detecting virial shocks in clusters of galaxies can be done using
simulated mock \ALMA\ observations and analyzing the simulated
visibilities with software packages used to analyze real data
(e.g. \MIRIAD, \AIPS). A detailed analysis would also make it
possible to find the optimal redshift for clusters to observe their
virial shocks.  We defer this detailed analysis to future work.

% % % % % % % % % % % % % % % % % % % % % % % % % % % % % % % % % % % % % % % % % % % % % % % % %
\subsection{Large Scale X-ray and SZ Signal Associated with Clusters of Galaxies}
\label{SS:LargeScaleSZX}

Soltan, Freyberg \& Hasinger (2002) have found evidence for large-scale extended X-ray 
emission out to about 2 degrees around rich Abell clusters using the \ROSAT\ All-Sky Survey. 
Assuming that this emission is originated from thermal bremsstrahlung within diffuse gas 
around clusters, Soltan et al.\ estimated the gas temperature to be below 1 keV, and 
concluded that this signal may originate from super-cluster gas -- gas left over from 
galaxy cluster formation and trapped by the gravitational potential well of super-clusters.

\cite{Myeret04} searched for the SZ decrement associated 
with clusters and groups of galaxies in the \WMAP\ data by cross-correlating it 
with the APM Galaxy Survey and the Abell-Corwin-Olowin catalog. 
They found evidence for an SZ signal around clusters of galaxies 
extending out to about 3\RVIR\ (assuming that most of the signal is originated from 
clusters with about $\RVIR \approx 1.75$ Mpc), 
and suggested that this signal is due to the hot super-cluster gas.
\cite{Afshet07} used WMAP 3 year data to stack images of 193 massive clusters of 
galaxies. Similar to the results of Myers et al., Afshordi et al. found that 
the distribution of the spherically averaged SZ signal of the stacked clusters 
extends out to at least about 3\RVIR\ (about 4 - 5 Mpc; see their Figure 2).

Most direct searches for super-cluster gas in individual objects -- as opposed to 
the stacked searches discussed in the preceding paragraph -- have not been successful 
\citep{Geno05,Boug99,MoBi98,Perset90}.
However, recently, \cite{Geno08} found evidence 
for SZ signal associated with the Corona Borealis super-cluster.
\cite{Zappet05} have found tentative evidence for diffuse, 
large scale emission from the Sculptor super-cluster. The estimated temperature of this
super-cluster was less then 0.5 keV.

Studies of the gas in superclusters can offer vital clues to the baryon evolution in strongly 
clustered environments.
Based on our cluster sample from the AMR simulations, we find that, in relaxed clusters, 
the external shocks are located at about 3\RVIR, and the region between the virial shocks and
the external shocks are filled with gas falling into the cluster.
Thus, our results suggest that a significant fraction of the large scale signal found around 
clusters of galaxies might be associated with infalling gas around individual massive clusters, 
and not with super-cluster gas 
(i.e. the signal is coming from gas bound to the individual massive clusters).

% % % % % % % % % % % % % % % % % % % % % % % % % % % % % % % % % % % % % % % % % % % % % % % % %
\subsection{Comparison with Other Results Based on Simulations}
\label{SS:Compare}

A comparison of our results to those based on other cosmological simulations is
 difficult, because we focus here on the high--resolution details of accretion shocks around 
 individual, massive clusters, whereas other studies focused on deriving global properties,
using a much larger simulation volume at lower resolution.
Nevertheless, we find that external shocks around massive clusters of galaxies form at about 
3\RVIR\ ($\pm 1\RVIR$), with large Mach numbers, extending to values as high as
$M = 10^4$.
Studies based on other simulations also found very high Mach numbers for external shocks
\citep{Miniet00,Ryu_et03,Keshet03,Pfroet06}.
Taking into account the photo-heating of the pre-shocked gas, \cite{Miniet00} found that the
Mach numbers of external shocks are less than a few hundred.
Based on the self-similar solution of \cite{Bert85}, scaled to our AMR simulations,
we estimate that external shocks at about 3\RVIR\ from the centers of massive 
clusters of galaxies should have Mach numbers of $\approx 90$ depending on 
the actual distance of the shock from cluster center.

Quantitative comparisons between our results for the AMR and SPH cluster samples
are hindered by the fact that the
AMR simulations were adiabatic, whereas the SPH simulations assumed that the gas was pre--heated.
The two simulations also used different cosmological parameters: $\sigma_8$=0.92 (AMR)
versus 1.0 (SPH), and $\Omega_b$=0.047 (AMR) versus 0.03 (SPH).
Overall, we find that the position of the virial shocks are much farther out in our 
SPH simulations (located at about 2-3\RVIR) 
than in the AMR simulations ($\approx 1$\RVIR); this is 
in spite of the fact that the SPH clusters have a lower baryon fraction.
This difference is probably not due to preheating: high--resolution 
AMR simulations of clusters of galaxies with and without preheating 
show no noticeable difference in the radial positions of the virial shocks 
(see Figure 1 of Younger \& Bryan 2007). 
Therefore, we attribute the large difference in the radial position of the virial shocks between our 
AMR and SPH simulations to the artificial viscosity and large smoothing
length far from the cluster center.
(We note that the problem may be indirectly exacerbated by the presence
of pre--heating, which keeps gas densities lower, and the smoothing length larger).
However, we expect the other features of virial shocks (such as the covering area and mass ratios)
in our SPH simulations to be reliable. Indeed, these quantities agree with
 those we found in the AMR cluster sample.

A comparison of the amount of mass locked up in filaments versus mass
in no--filament regions is also of interest, because it tells us
roughly how much mass will be accreted via filaments in the form of
dense clumps, collapsed halos and gas collapsed into filaments versus
via accretion of low--density diffuse gas. We determined this mass ratio in 
regions outside of the smoothed virial radius and thus should be free from 
the numerical problems mentioned in the previous paragraph.  
We show the total mass in filaments as a percentage of the total mass for 
all clusters in our AMR and SPH cluster sample in Tables 1 and 2.  
About 80\% to 90\% of the mass resides in the filaments falling toward the
cluster centers both in our AMR and SPH sample.  The slightly larger
percentage of mass in filaments in our SPH simulations is most likely
due to the higher value of $\sigma_8$ used in those simulations.
Based on our simulations, we conclude that massive clusters of
galaxies accrete most of their mass via filaments, rather than via
low--density diffuse gas, even at the later stages of their evolution.

% % % % % % % % % % % % % % % % % % % % % % % % % % % % % % % % % % % % % % % % % % % % % % % % %
\section{Conclusion}
\label{S:Conclusion}

We have studied large-scale accretion shocks in a sample of massive
($\approx 10^{15}~\MSUN$) galaxy clusters at redshift $z=0$ drawn from
AMR and SPH cosmological \LCDM\ simulations.  The properties of shocks
in clusters with a larger range of masses and redshifts will be
studied in a forthcoming paper.  Most of our quantitative results are
based on the cluster sample drawn from our AMR simulations, which did
not take feedback (heating, cooling, star--formation, and supernovae)
into account.  However, the properties of the shocks we derived from
these simulations agree with those in an independent set of SPH
simulations, with the exception of the radial location of the virial
shocks (which are found to be farther out in the SPH runs, due to
numerical effects).  

The main results of this study can be summarized as follows.
(1) Clusters are surrounded by two distinct sets of accretion shocks,
located near $\sim 1\RVIR$, and $\sim 3\RVIR$, respectively.
(2) These shocks cover $\sim 50\%$ of the surface area around each
cluster.
(3) The shocks near $\sim 1\RVIR$ cause a sharp drop in
the projected SZ surface brightness, tracing out a circular
arc in projection over a non--negligible fraction (10--50\%; depending
strongly on both the dynamical state of the cluster, and on the viewing angle) of the 2$\pi$
azimuthal angle around the cluster center.  
(4) The significance of the SZ feature due to these shocks is reduced by a factor
of 5-10 relative to earlier predictions based on spherical models \citep{Kocset05},
because of the partial extent of the sharp arc, and because the simulations
predict steeper SZ surface brightness profiles.  Nevertheless, the features may 
be detectable with dedicated mosaic observations, by
 the next generation high angular resolution and sensitivity radio interferometer, \ALMA.

Our results should motivate a more detailed follow--up study of the
effects of non-spherical accretion, astrophysical and instrumental
noise, and image processing techniques on the detectability of virial
shocks in massive clusters of galaxies using realistic mock \ALMA\
observations of clusters drawn from high--resolution cosmological
simulations.  Such studies will be necessary for a more robust
estimate of the feasibility of the detection of these shocks, as well
as to find the optimal redshift for the observations, and to determine
the best target selection method and observation strategy.

% % % % % % % % % % % % % % % % % % % % % % % % % % % % % % % % % % % % % % % % % % % % % % % % %
% 
% ACKNOWLEDGMENTS
% 
% % % % % % % % % % % % % % % % % % % % % % % % % % % % % % % % % % % % % % % % % % % % % % % % %
\acknowledgments

We thank M. Birkinshaw, F. Miniati, and P. Ricker for useful
discussions, and the anonymous referee for comments that significantly
improved this manuscript.  We acknowledge the use of \HEALPIX\ (Gorski
et al., 2005; http://healpix.jpl.nasa.gov).  AEE acknowledges support
from NSF grant AST-0708150 and NASA ATP Grant NAG5-13378.  GB
acknowledges support from NSF grants AST-05-07161, AST-05-47823 and
supercomputing resources from the National Center for Supercomputing
Applications.  ZH was supported by the NSF grant AST-05-07161 and by
the Pol\'anyi Program of the Hungarian National Office for Research
and Technology (NKTH).

\newpage
% % % % % % % % % % % % % % % % % % % % % % % % % % % % % % % % % % % % % % % % % % % % % % % % %
%
%                                      B I B L I O G R A P H Y
%
% % % % % % % % % % % % % % % % % % % % % % % % % % % % % % % % % % % % % % % % % % % % % % % % %
\bibliographystyle{apj}

\newpage
% % % % % % % % % % % % % % % % % % % % % % % % % % % % % % % % % % % % % % % % % % % % % % % % %
% % % % % % % % % % % % % % % % % % % % % % % % % % % % % % % % % % % % % % % % % % % % % % % % %
% % % % % % % % % % % % % % % % % % % % % % % % % % % % % % % % % % % % % % % % % % % % % % % % %
%
%                                      F I G U R E S
%
% % % % % % % % % % % % % % % % % % % % % % % % % % % % % % % % % % % % % % % % % % % % % % % % %
% % % % % % % % % % % % % % % % % % % % % % % % % % % % % % % % % % % % % % % % % % % % % % % % %
% % % % % % % % % % % % % % % % % % % % % % % % % % % % % % % % % % % % % % % % % % % % % % % % %

% % % % % % % % % % % % % % % % % % % % % % % % % % % % % % % % % % % % % % % % % % % % % % % % %
%  
%  FIGURE 1
%  
% % % % % % % % % % % % % % % % % % % % % % % % % % % % % % % % % % % % % % % % % % % % % % % % %
 % Figure~\ref{F:Fig1}

\clearpage

% % % % % % % % % % % % % % % % % % % % % % % % % % % % % % % % % % % % % % % % % % % % % % % % %
%  
%  FIGURE 2
%  
% % % % % % % % % % % % % % % % % % % % % % % % % % % % % % % % % % % % % % % % % % % % % % % % %
 % Figure~\ref{F:Fig2}

\clearpage

% % % % % % % % % % % % % % % % % % % % % % % % % % % % % % % % % % % % % % % % % % % % % % % % %
%  
%  FIGURE 3
%  
% % % % % % % % % % % % % % % % % % % % % % % % % % % % % % % % % % % % % % % % % % % % % % % % %
 % Figure~\ref{F:Fig3}

\clearpage

% % % % % % % % % % % % % % % % % % % % % % % % % % % % % % % % % % % % % % % % % % % % % % % % %
%  
%    FIGURE 4
%  
% % % % % % % % % % % % % % % % % % % % % % % % % % % % % % % % % % % % % % % % % % % % % % % % %

% % % % % % % % % % % % % % % % % % % % % % % % % % % % % % % % % % % % % % % % % % % % % % % % %
% VIRIAL SHOCKS: RSHOCK AND MACH NUMBER DISTRIBUTION
% % % % % % % % % % % % % % % % % % % % % % % % % % % % % % % % % % % % % % % % % % % % % % % % %
                 %  Figure~\ref{F:Fig4}

\clearpage

% % % % % % % % % % % % % % % % % % % % % % % % % % % % % % % % % % % % % % % % % % % % % % % % %
% 
%  FIGURE 5
% 
% % % % % % % % % % % % % % % % % % % % % % % % % % % % % % % % % % % % % % % % % % % % % % % % %
 % Figure~\ref{F:Fig5}

\clearpage

% % % % % % % % % % % % % % % % % % % % % % % % % % % % % % % % % % % % % % % % % % % % % % % % %
% 
%  FIGURE 6
% 
% % % % % % % % % % % % % % % % % % % % % % % % % % % % % % % % % % % % % % % % % % % % % % % % %
 % Figure~\ref{F:Fig6}

\clearpage

% % % % % % % % % % % % % % % % % % % % % % % % % % % % % % % % % % % % % % % % % % % % % % % % %
%  
%  FIGURE 7
%  
% % % % % % % % % % % % % % % % % % % % % % % % % % % % % % % % % % % % % % % % % % % % % % % % %
 % Figure~\ref{F:Fig7}

\clearpage

% % % % % % % % % % % % % % % % % % % % % % % % % % % % % % % % % % % % % % % % % % % % % % % % %
%  
%  FIGURE 8
%  
% % % % % % % % % % % % % % % % % % % % % % % % % % % % % % % % % % % % % % % % % % % % % % % % %
 % Figure~\ref{F:Fig8}

\clearpage

% % % % % % % % % % % % % % % % % % % % % % % % % % % % % % % % % % % % % % % % % % % % % % % % %
% 
% FIGURE 9
% 
% % % % % % % % % % % % % % % % % % % % % % % % % % % % % % % % % % % % % % % % % % % % % % % % %
 % Figure~\ref{F:Fig9}

\clearpage

% % % % % % % % % % % % % % % % % % % % % % % % % % % % % % % % % % % % % % % % % % % % % % % % %
%  
%  FIGURE 10
% 
% % % % % % % % % % % % % % % % % % % % % % % % % % % % % % % % % % % % % % % % % % % % % % % % %
 % Figure~\ref{F:Fig10}

\clearpage

% % % % % % % % % % % % % % % % % % % % % % % % % % % % % % % % % % % % % % % % % % % % % % % % %
% 
%  FIGURE 11
% 
% % % % % % % % % % % % % % % % % % % % % % % % % % % % % % % % % % % % % % % % % % % % % % % % %

\clearpage

% % % % % % % % % % % % % % % % % % % % % % % % % % % % % % % % % % % % % % % % % % % % % % % % %
% % % % % % % % % % % % % % % % % % % % % % % % % % % % % % % % % % % % % % % % % % % % % % % % %
%  
%  TABLES
%  
% % % % % % % % % % % % % % % % % % % % % % % % % % % % % % % % % % % % % % % % % % % % % % % % %
% % % % % % % % % % % % % % % % % % % % % % % % % % % % % % % % % % % % % % % % % % % % % % % % %

% % % % % % % % % % % % % % % % % % % % % % % % % % % % % % % % % % % % % % % % % % % % % % % % %
% 
% TABLE 1 : AMR CLUSTERS
% 
% % % % % % % % % % % % % % % % % % % % % % % % % % % % % % % % % % % % % % % % % % % % % % % % %
\begin{deluxetable}{ccccccccccc}
\tablecolumns{11}
\tablecaption{
\label{tab:Table1}
 AMR Cluster Parameters} 
\tablewidth{0pt} 
\tablehead{ 
 \multicolumn{1}{c}{CL ID\tablenotemark{a}} &
 \multicolumn{1}{c}{{$\rm M_{vir}$}\tablenotemark{b}} &
 \multicolumn{1}{c}{{$\rm R_{vir}$}\tablenotemark{c}} &
 \multicolumn{1}{c}{Dyn.State\tablenotemark{d}} &
 \multicolumn{1}{c}{{$\rm {\cal R}_{vsh}$}\tablenotemark{e}} &
 \multicolumn{1}{c}{{$\rm {\cal M}_{vsh}$}\tablenotemark{f}} &
 \multicolumn{1}{c}{{$\rm A_{vsh}$}\tablenotemark{g}} &
 \multicolumn{1}{c}{$\rm {\cal R}_x$\tablenotemark{h}} &
 \multicolumn{1}{c}{$\rm {\cal M}_x$\tablenotemark{i}} &
 \multicolumn{1}{c}{$\rm A_x$\tablenotemark{j} } &
 \multicolumn{1}{c}{{$\rm M_{fil}/M_{tot}$}\tablenotemark{k}}
}
\startdata  
AMR1  &   2.3E+15  &   2.7  &  Merging  &   0.9  &   5.1  &    37\%  &   2.2  &    90.1  &    49\%  &    76\%   \\
AMR2  &   2.0E+15  &   2.6  &  Relaxed  &   0.9  &   2.6  &    37\%  &   3.2  &   112.5  &    52\%  &    58\%   \\
AMR3  &   1.5E+15  &   2.4  &  Merging  &   1.2  &   5.0  &    25\%  &   2.3  &   109.3  &    45\%  &    79\%   \\
AMR4  &   1.4E+15  &   2.3  &  Merging  &   1.2  &   3.2  &    23\%  &   2.9  &   179.7  &    46\%  &    80\%   \\
AMR5  &   1.2E+15  &   2.2  &  Relaxed  &   1.0  &   3.3  &    43\%  &   2.8  &    91.3  &    58\%  &    75\%   \\
AMR6  &   1.1E+15  &   2.1  &  Merging  &   1.4  &   2.7  &    18\%  &   2.7  &    95.0  &    61\%  &    78\%   \\
AMR7  &   1.1E+15  &   2.1  &  Merging  &   1.2  &   2.6  &    24\%  &   3.5  &   242.4  &    44\%  &    82\%   \\
AMR8  &   1.1E+15  &   2.1  &  Merging  &   1.0  &   3.3  &    23\%  &   2.2  &   139.1  &    30\%  &    90\%   \\
AMR9  &   1.1E+15  &   2.1  &  Relaxed  &   1.3  &   4.0  &    46\%  &   2.7  &    87.5  &    36\%  &    74\%   \\
AMR10  &   9.1E+14  &   2.0  &  Merging  &   1.4  &   3.1  &    28\%  &   3.6  &   135.2  &    44\%  &    90\%
\enddata
\tablenotetext{a}{Cluster ID.}
\tablenotetext{b}{Virial mass (in \MSUN).}
\tablenotetext{c}{Virial radius (in Mpc).}
\tablenotetext{d}{Dynamical state.}
\tablenotetext{e}{Characteristic radial position of the virial shock in units of $\rm R_{vir}$.}
\tablenotetext{f}{Characteristic Mach number of the virial shock.}
\tablenotetext{g}{Area covered by virial shock (percentage of the total solid angle, 4$\pi$).}
\tablenotetext{h}{Radial position of the external shock in units of $\rm R_{vir}$.}
\tablenotetext{i}{Characteristic Mach number of the external shock.}
\tablenotetext{j}{Surface area covered by external shocks (percentage of the total area).}
\tablenotetext{k}{Mass in filaments over total mass in the infall region 
                           ($\rm M_{fil}/M_{tot}$ expressed as percentage).}
\label{T:Table1}
\end{deluxetable} % Table~\ref{T:Table1}

% % % % % % % % % % % % % % % % % % % % % % % % % % % % % % % % % % % % % % % % % % % % % % % % %
% 
% TABLE 2 : SPH
% 
% % % % % % % % % % % % % % % % % % % % % % % % % % % % % % % % % % % % % % % % % % % % % % % % %
\begin{deluxetable}{ccccccc}
\tablecolumns{7}
\tablecaption{
 SPH Cluster Parameters} 
\tablewidth{0pt} 
\tablehead{ 
 \multicolumn{1}{c}{CL ID\tablenotemark{a}} &
 \multicolumn{1}{c}{{$\rm M_{vir}$}\tablenotemark{b}} &
 \multicolumn{1}{c}{{$\rm M_{vir}$}\tablenotemark{c}} &
 \multicolumn{1}{c}{Dyn.State\tablenotemark{d}} &
 \multicolumn{1}{c}{{$\rm {\cal R}_{vsh}$}\tablenotemark{e}} &
 \multicolumn{1}{c}{{$\rm A_{vsh}$}\tablenotemark{f}} &
 \multicolumn{1}{c}{{$\rm M_{fil}/M_{tot}$}\tablenotemark{g}}
}
\startdata  
SPH1  &   2.1E+15  &   2.7  &  Relaxed  &   2.3  &    56\%   &    94\%   \\
SPH2  &   1.7E+15  &   2.5  &  Merging  &   2.1  &    43\%   &    41\%   \\
SPH3  &   1.7E+15  &   2.5  &  Merging  &   3.1  &    19\%   &    87\%   \\
SPH4  &   1.4E+15  &   2.4  &  Relaxed  &   2.7  &    45\%   &    83\%   \\
SPH5  &   1.4E+15  &   2.4  &  Merging  &   2.2  &    35\%   &    91\%   \\
SPH6  &   1.4E+15  &   2.4  &  Relaxed  &   1.9  &    36\%   &    90\%   \\
SPH7  &   1.3E+15  &   2.3  &  Merging  &   2.1  &    40\%   &    84\%   \\
SPH8  &   1.2E+15  &   2.3  &  Merging  &   2.9  &    42\%   &    91\%   \\
SPH9  &   1.2E+15  &   2.2  &  Merging  &   2.7  &    44\%   &    92\%   \\
SPH10  &   1.1E+15  &   2.2  &  Relaxed  &   2.4  &    45\%   &    92\%
\enddata
\tablenotetext{a}{Cluster ID.}
\tablenotetext{b}{Virial mass (in \MSUN).}
\tablenotetext{c}{Virial radius (in Mpc).}
\tablenotetext{d}{Dynamical state.}
\tablenotetext{e}{Characteristic radial position of the virial shock in units of $\rm R_{vir}$.}
\tablenotetext{f}{Surface area covered by virial shocks (percentage of the total area).}
\tablenotetext{g}{Mass in filaments over total mass in the infall region 
                           ($\rm M_{fil}/M_{tot}$, expressed as percentage).}
\label{T:Table2}
\end{deluxetable} % Table~\ref{T:Table2}

% % % % % % % % % % % % % % % % % % % % % % % % % % % % % % % % % % % % % % % % % % % % % % % % %
 
\end{document}